\documentclass[twocolumn]{aastex631}
\usepackage{amsmath,amsfonts,amssymb,graphicx,chngcntr,multirow, float,booktabs, afterpage}
\usepackage[]{hyperref}
\hypersetup{colorlinks=true}

\usepackage{longtable}

\received{}
\revised{}
\accepted{}

\shorttitle{Weighing SMBHs with TDE optical flares }

\begin{document}

\title{Supermassive black hole mass inference with the optical flares of tidal disruption events}

\author{A. Mummery}
\affiliation{School of Natural Sciences, Institute for Advanced Study, 1 Einstein Drive, Princeton, NJ 08540, USA}

\begin{abstract}
 Tidal disruption events (TDEs) represent a truly unique, and potentially very powerful,  probe of the quiescent supermassive black hole (SMBH) population. Given current observational survey capabilities the vast majority of the TDEs discovered in the next decade will be observed only across optical-UV wavelengths. A set of questions of broad scientific interest relating to SMBH demographics and SMBH-galaxy correlations could in principal be answered by using TDE emission as an efficient means to constrain SMBH masses. In this paper we argue for using well-understood elements of TDE emission (the thermal X-ray continuum and late-time UV plateau) to derive empirical relationships between the more poorly understood early optical/UV flare and the black hole mass, before using these empirical relationships to measure TDE black hole masses simply and rapidly. We provide a publicly available code {\tt TDEFLARE} which does this, showing (i) it produces results consistent with disk codes containing far more physics, (ii) it reproduces galactic scaling relationships at high ($>5\sigma$) significance, (iii) it produces reliable mass estimates for both partial and full disruptions, and (iv) it does not require late time data to derive mass constraints. We provide 89 TDE black hole mass constraints,  derive the intrinsic black hole mass function implied by the current TDE population, and discuss the Malmquist-Hills bias, an important confounding factor in TDE science. 
\end{abstract}


\section{Introduction} \label{sec:intro}
The tidal disruption and subsequent accretion of a star by the supermassive black hole at the center of a previously quiescent galaxy powers multi wavelength flares now routinely detected across the electromagnetic spectrum. Predominantly detected in the X-ray and across optical/UV frequencies, the {\it intrinsic} rate of optically bright and X-ray bright TDEs is comparable \citep[e.g.][]{Guolo24, Sazonov21, Yao23, Grotova25}, but observational capabilities (i.e., survey efficiency not  intrinsic properties) mean that optical discoveries dominate the TDE population numerically. This simple fact is only going to become more consequential as we enter the Rubin/LSST era \citep{Bricman20}, whereafter the number of (optically bright) TDEs is anticipated to grow by orders of magnitude (from the present population of $\sim 100$).  

Tidal disruption events offer a promising avenue to answer a number of questions of fundamental importance in black hole astrophysics: what is the occupation fraction of intermediate mass black holes in dwarf galaxies? Do galactic scaling relationships hold down to low masses? What is the distribution of supermassive black hole masses in the local universe? 

A number of the most important astrophysical questions which can be probed by TDE science therefore relate to the masses of the black holes in the center of these events (and their distributions across a population). Owing to the dominance of optically-bright sources in a TDE sample (an inevitable result of their  survey efficiencies being highest), it is clear that ultimately being able to relate optical/UV features of TDE emission to black hole masses is an essential step in answering these fundamental questions.  

If one has long (multi-year) timescale well-sampled data from a nearby source, this is a solved problem. After the initial flare  (i.e., at late times, $\Delta t \gtrsim 1\, {\rm yr}$) the optical/UV emission from a TDE transitions to a plateau (weakly time dependent) phase \citep[][]{vanVelzen19, MumBalb20a, Mummery_et_al_2024}, which  arises naturally from a spreading disk formed from the stellar debris (this disk also produces the soft X-ray observed at all times, e.g., \citealt{MummeryVV25,Guolo25c}).   Importantly, modeling this late time emission within this framework enables an estimate of the black hole mass at the center to be computed, which is known to agree with the $M_\bullet-\sigma$ value \citep{Mummery_et_al_2024} and the $M_\bullet-M_{\rm bulge}$ value \citep{Ramsden25}. Even better (in the sense that they are more precise) black hole mass inference results if one also incorporates X-ray emission at these late stages of evolution \citep{Guolo25c}. Every TDE which is nearby $z<0.08$ and has $>2$ years of data has a detected plateau \citep[][for $N = 40$ sources]{Mummery25_cal}, meaning that this disk formation is ubiquitous, and can always be used when the source is sufficiently nearby for this late time emission to be detected. 

However, securing this late time data is expensive -- one has to wait a long time post TDE discovery, one must stack many epochs of data together to get a robust detection, and (in  general)  one needs UV data for the best constraints, something which is not routinely detected with optical surveys. In an ideal world, one would be able to perform black hole mass inference rapidly, with the relatively small amount of data taken near peak. This will also be essential for the many high redshift TDEs discovered by Rubin/LSST, which may well be too distant for a robust detection of the plateau. 

Many models of the early optical/UV flare detected in TDEs exist, including stream-stream collisions \citep[e.g.,][]{Dai+15,Ryu+20a}, the efficient reprocessing of the fallback rate \citep{Mockler19}, the reprocessing of black hole accretion by accretion disk winds \citep[e.g.,][]{Strubbe&Quataert09,MetzgerStone16}, or the gravitational contraction of an extended quasi-spherical ``envelope'' \citep{Metzger22}. Models based on shocks \citep{Ryu+20a} and fallback reprocessing \citep{Mockler19} have been routinely used in the community, and fail to recover known galactic scaling relationships for the black hole mass at a statistically significant level \citep{Hammerstein23, Ramsden22, Guolo25c}. This is likely because they make assumptions about the scaling of the luminosity of the flare with black hole mass which are ruled out by the data at extremely high significance ($>5\sigma$; \citealt{Mummery25_cal}, \citealt{Guolo25c}). Other models make assumptions which are in much better accord with the data \citep[e.g., the cooling envelope model of][]{Metzger22} but are currently incomplete in terms of their utility at all black hole mass scales and must be extended (see extended discussion in \citealt{Mummery25_cal}). 

The underlying physics governing the optical flares in TDE is undoubtedly complicated, and represents a fundamentally hard theoretical problem. It is perhaps surprising therefore that in a practical sense, recovering the gross properties of these TDE flares empirically is a simple (and solved) problem. Using black hole masses calibrated from accretion disk models, \cite{Mummery_et_al_2024} showed that the peak luminosity and radiated energy in a TDE's optical flare is tightly correlated with the black hole mass in the center. With a black hole mass function constructed from first principles accretion disk theory and late time UV data, one gets the peak optical luminosity function ``for free'' by using these empirically calibrated relationships \citep{MummeryVV25} suggesting that there are minimal compounding variables that need be tracked beyond the black hole mass. The question is then how to best utilize these scaling relationships for black hole mass inference. 

In this paper we present the code ``Tidal Disruption Event Fast Likelihood AccRetion Estimate ({\tt TDEFLARE})''\footnote{\href{https://bitbucket.org/fittingtransientswithdiscs/fitted_public/src}{https://bitbucket.org/fittingtransientswithdiscs/fitted\_public/src}} which is a code designed to satisfy the observational constraints relevant for the Rubin/LSST era. It is designed to be (i) rapid to fit to data, typically taking seconds on a laptop (as opposed to the much longer times for more physically motivated disk codes), (ii) provide tight mass constraints in the presence {\it or absence} of a detected plateau (either because this is early in the TDE evolution or because the late-time emission is too faint to detect), (iii) uses {\it empirical} relationships {\it calibrated with well understood disk physics} to model the parts of the optical/UV flare which are more poorly understood. 

In effect this framework treats the optical/UV flare of TDEs much like type 1a supernova are treated in cosmology -- we do not fully understand all of the physics of the event, but we know there is an empirical way to constrain the parameter we are after and this can be useful for answering important questions. 

We show that this framework results in reliable (in the sense that they reproduce known scaling relationships) black hole mass inference for full and (repeating) partial TDEs. The model is quick to fit to data, and produces mass posteriors which are consistent with much more physically motivated models (e.g., {\tt FitTeD} \citealt{mummery2024fitted} and {\tt kerrSED} \citealt{GuoloMum24}). This model produces  black hole mass posteriors which are consistent with these more complex disk models using only the data taken near the peak of the flare, but performs best when late time data is available.  

The layout of this paper is the following. In section \ref{method} we introduce the scaling relationships used, the statistical techniques employed, and the code structure. In section \ref{implementation} we show some example fits to three TDEs, chosen to represent a broad class of objects. We then compare the results of {\tt TDEFLARE} to other, significantly more physically motivated disk codes, finding excellent agreement. In section \ref{discussion} we use {\tt TDEFLARE} on the entire TDE population, and discuss various results and biases. We conclude in section \ref{conclusions}. Black hole mass constraints of the modeled TDEs are listed in the Appendix. 

\section{Method}\label{method}
\subsection{Scaling relationships}
The {\tt TDEFLARE} model is built around three scaling relationships between the black hole mass in a TDE $(M_\bullet)$ and properties of their optical/UV lightcurves. These relationships were published in \cite{Mummery_et_al_2024}, and relate the ``plateau'' luminosity $L_P$, the ``peak'' luminosity ($L_{\rm pk}$) and the ``energy radiated'' ($E_g$) to the mass of the central black hole (all three observables will be defined precisely shortly). 

The plateau luminosity-black hole mass scaling relationship is derived from first-principles time-dependent relativistic accretion theory, and takes the following form
\begin{equation}
    \log_{10} \left({M_\bullet \over M_\odot}\right) = 1.50 \, \log_{10} \left({ L_{P} \over 10^{41} \, {\rm erg}\,{\rm s}^{-1}}\right) + 6.0  , 
\end{equation}
where $L_{P}$ is the late-time value of $\nu L_\nu$ measured in the sources rest-frame $g$-band $(\nu = 6 \times10^{14} \, {\rm Hz})$. We do not quote an uncertainty on this scaling relationship, because it is ``theoretical''. We will use an empirical measure of the scatter in this relationship when inferring black hole masses, however. It is important to stress the (potentially surprising) result proved in Mummery (in prep.), that this scaling relationship holds for both full {\it and partial} TDEs, with a very weak impact parameter dependence. This is an important result, as the fraction of observed TDEs which are actually partial TDEs is poorly constrained, but this uncertainty will not impact mass inference derived via the plateau luminosity. 

The above scaling relationship provides a first-principles method of measuring black hole masses in TDEs. These masses are consitent with the $M_\bullet-\sigma$, $M_\bullet-M_{\rm gal}$ and $M_\bullet-M_{\rm bulge}$ scaling relationships \citep{Mummery_et_al_2024, MummeryVV25, Ramsden25}. This  result was utilized by \cite{Mummery_et_al_2024} to calibrate empirical (i.e., containing no intrinsic fundamental physics) scaling relationships between the ``peak'' luminosity ($L_{\rm pk}$) and the ``energy radiated'' ($E_g$) in the sources $g$-band, with the mass measured from the plateau $M_\bullet(L_P)$. 

The peak $g$-band luminosity-black hole mass scaling relationship takes the form 
\begin{multline}
    \log_{10} \left({M_\bullet \over M_\odot}\right) = (0.98 \pm 0.10) \, \log_{10} \left({ L_{\rm pk} \over 10^{43} \, {\rm erg}\,{\rm s}^{-1}}\right) \\ + (6.52 \pm 0.06)  , 
\end{multline}
where $L_{\rm pk}$ is again $\nu L_\nu$ measured in the sources rest-frame $g$-band $(\nu = 6 \times10^{14} \, {\rm Hz})$, but now measured at the peak of the observed emission. The quoted uncertainty on this scaling relationship represents $68\%$ (i.e., $1\sigma$) posterior confidence interval. However, the scatter in this relationship is dominated by the intrinsic scatter, not the uncertainty in the indices. 

Similarly, the radiated $g$-band energy-black hole mass scaling relationship takes the form 
\begin{multline}
    \log_{10} \left({M_\bullet \over M_\odot}\right) = (0.98 \pm 0.07) \, \log_{10} \left({ E_{g} \over 10^{50} \, {\rm erg}}\right) \\ + (6.78 \pm 0.04)  , 
\end{multline}
where $E_{g}$ is also measured in the sources rest-frame $g$-band $(\nu = 6 \times10^{14} \, {\rm Hz})$, and we define the radiated energy as the integrated energy in the decaying portion of the initial TDE flare (as rise data is not always available). Again, the quoted uncertainty on this scaling relationship represents $68\%$ (i.e., $1\sigma$) posterior confidence interval, and the scatter in this relationship is again dominated by  intrinsic scatter.

The intrinsic scatters in these relationships was measured by \cite{Mummery_et_al_2024, Mummery25_cal}, and have the following values in log space 
\begin{align}
    \epsilon_{L_{\rm pk}} &= 0.53\pm 0.04\,\, {\rm dex}, \\
    \epsilon_{E_{g}} &= 0.44\pm 0.03\,\, {\rm dex}, \\
    \epsilon_{L_P} &= 0.38 \pm 0.04\,\, {\rm dex}.
\end{align}

\begin{figure*}
    \centering
    \includegraphics[width=0.45\linewidth]{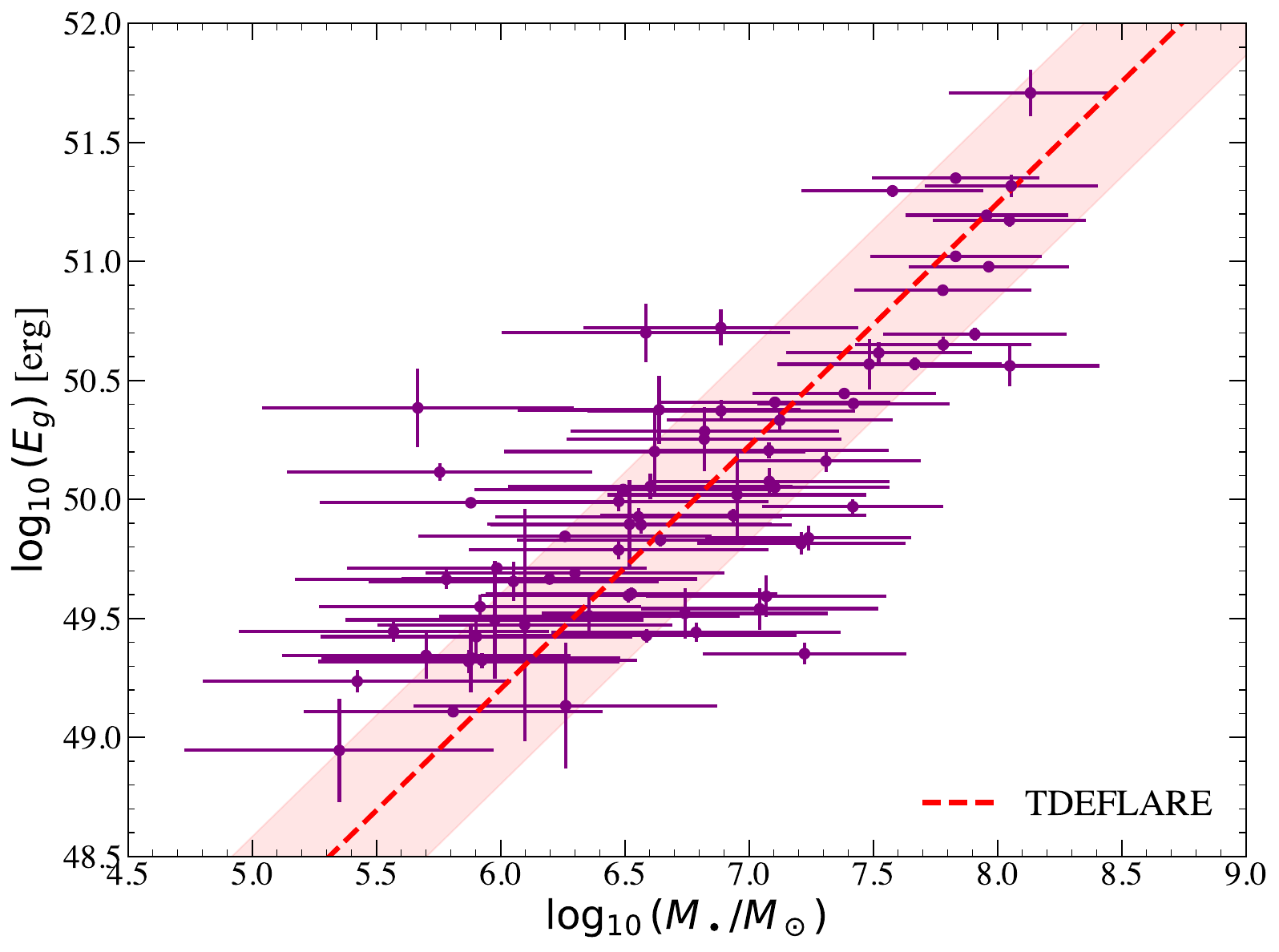}
    \includegraphics[width=0.45\linewidth]{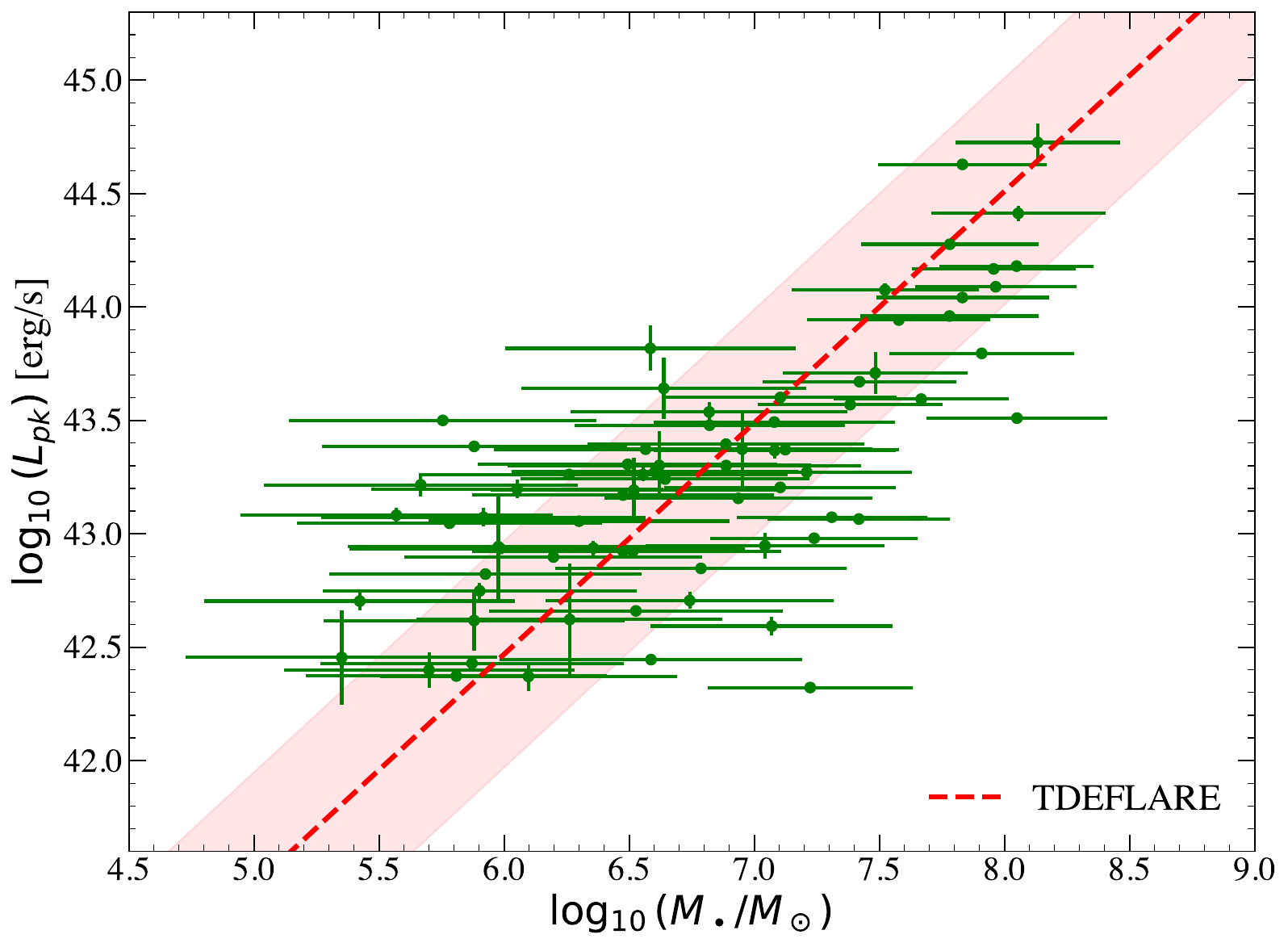}
    \includegraphics[width=0.67\linewidth]{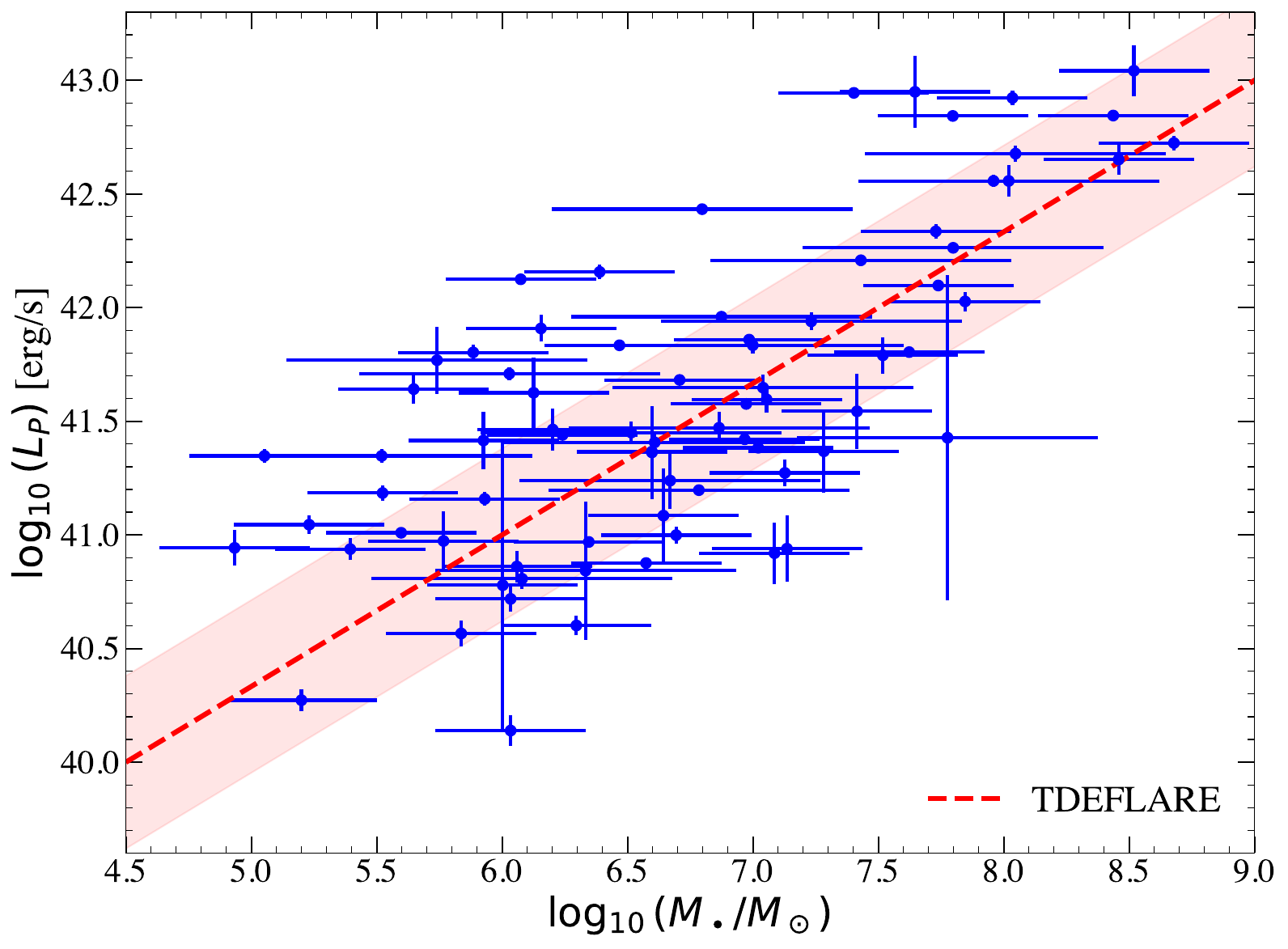}
    \caption{The scaling relationships employed by {\tt TDEFLARE} between the black hole mass and the radiated $g$-band energy (upper left), the peak $g$-band luminosity (upper right) and the $g$-band plateau luminosity (lower). The lower panel plots black hole masses as inferred from galactic scaling relationships ($M_\bullet-\sigma$ where available, and if not $M_\bullet-M_{\rm bulge}$ where available, and otherwise $M_\bullet-M_{\rm gal}$) against observed plateau luminosities, while the curve and shaded region is derived from first principles disk theory. The upper two panels use the disk-theory informed black hole masses to calibrate empirical scaling relationships between the radiated energy/peak luminosity and the black hole mass. These curves are not based on a fundamental theory, but remain of practical use in constraining black hole masses from TDE data. In each plot the dashed curve represents the median of the scaling relationship, while the shaded region denotes the width of one $\epsilon_i$ scatter (in log space), for each of the three scaling relationships (see text). }
    \label{fig:basis}
\end{figure*}

\begin{figure*}
    \centering
    \includegraphics[width=0.48\linewidth]{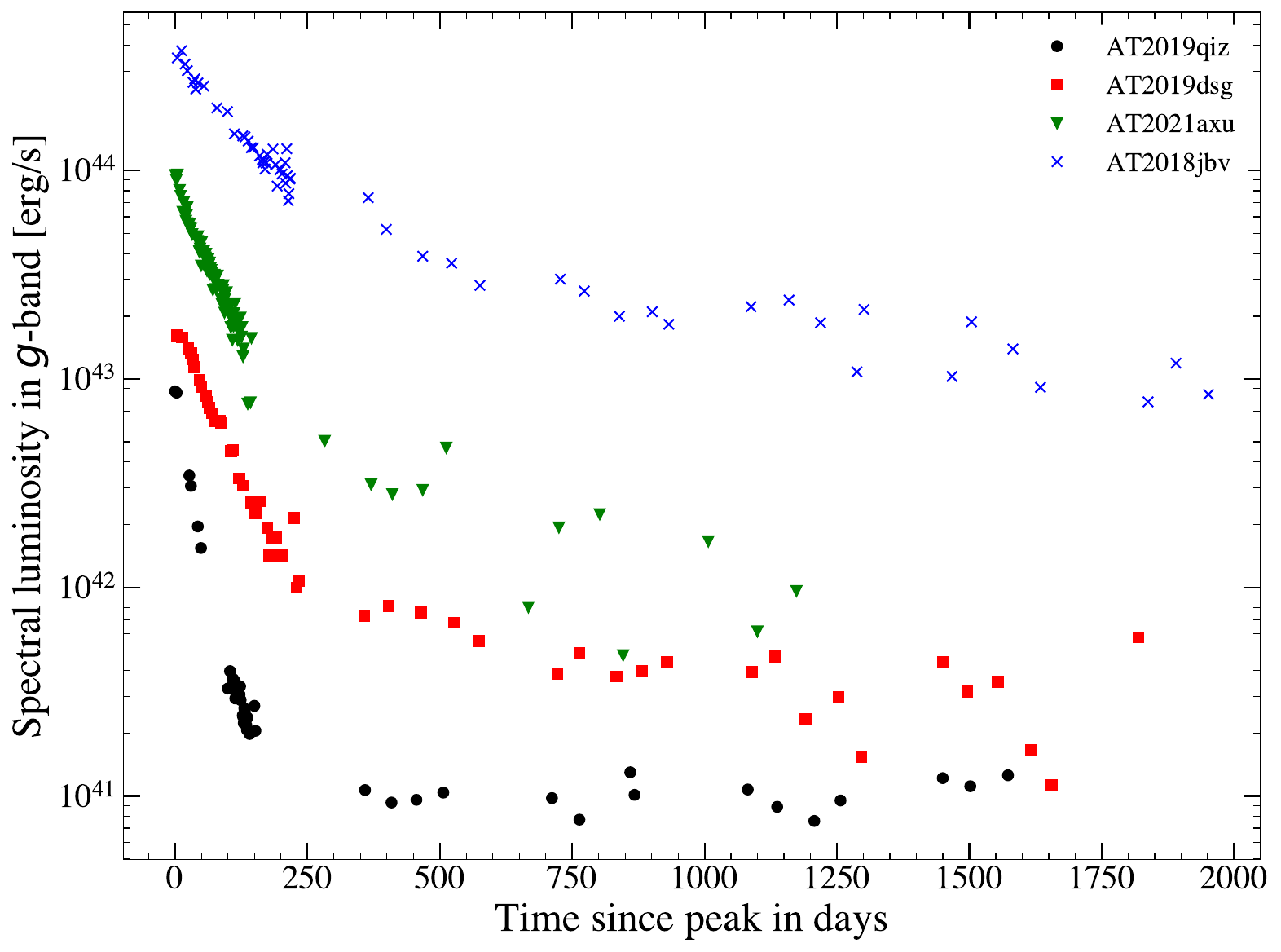}
    \includegraphics[width=0.48\linewidth]{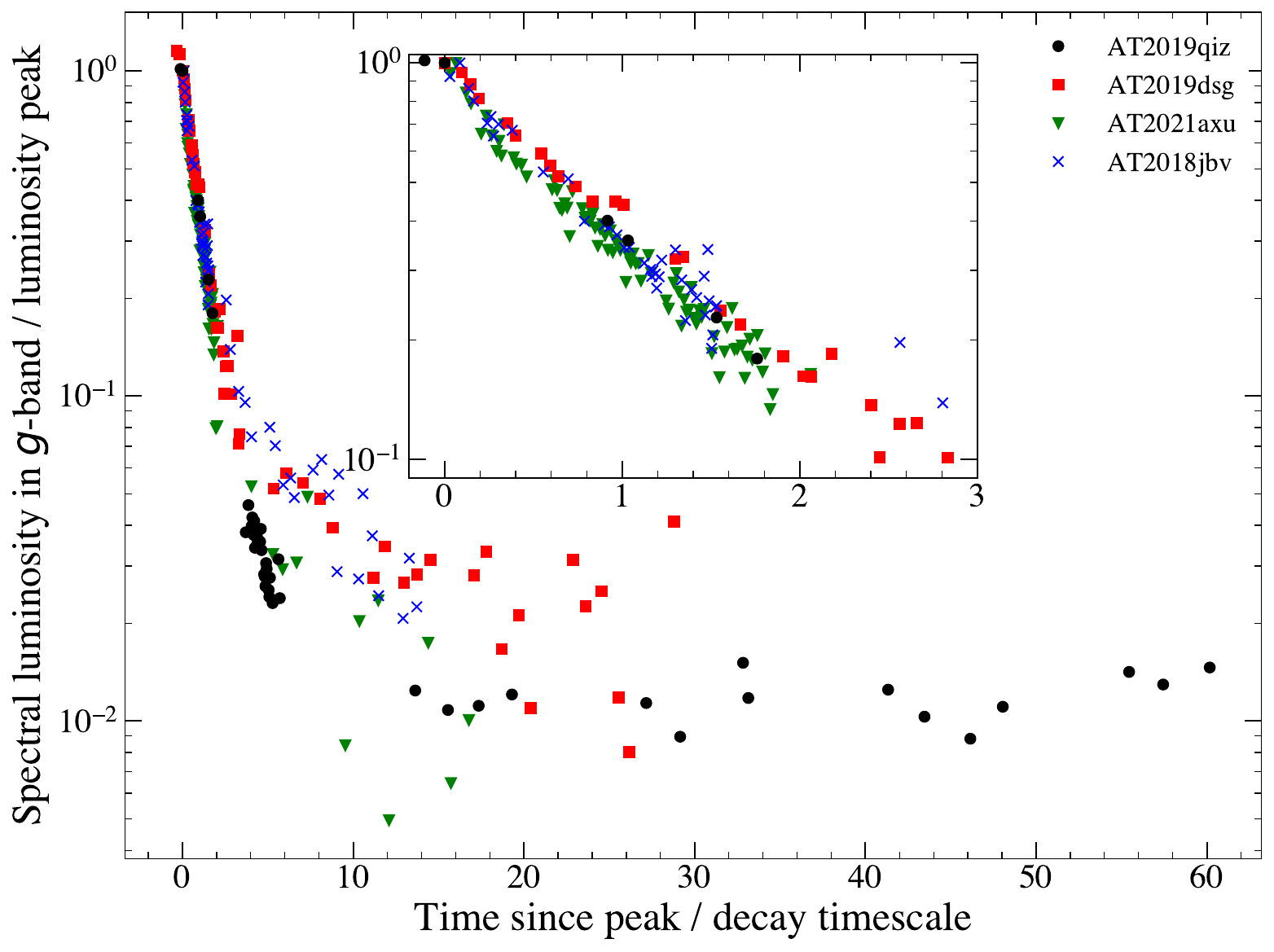}
    \caption{The optical flares from TDEs are, broadly speaking, simple to describe over their first decade of luminosity decay in the sense that they are well described by exponential profiles. Left: the optical ($g$-band) light curves of four TDEs (names on plot), chosen to span the full range of peak luminosities seen in TDEs. Right: these light curves ``normalised'', i.e., the observed luminosity normalised by their peak spectral luminosity, and with time axis normalised by a simple exponential decay timescale $\tau$ (fit to each TDEs first $\sim 100$ days of $g$-band data). The inset shows the first decade of decay, which by definition takes $\ln(10)\approx 2.3$ decay times. Brighter TDEs decay more slowly (the opposite of what would be expected from fallback arguments). This implies that the initial flare from TDEs can be described by two numbers in an exponential profile (with minimal ``hidden'' variables), both of which can be related to the black hole mass in the system.  }
    \label{fig:standard}
\end{figure*}

We plot these three scaling relationships in Figure \ref{fig:basis}. The lower panel shows the black hole masses as inferred from galactic scaling relationships ($M_\bullet-\sigma$ where available, and if not $M_\bullet-M_{\rm bulge}$ where available, and otherwise $M_\bullet-M_{\rm gal}$) against the observed plateau luminosities \citep[data from][for the plateau luminosities and $\sigma$ and $M_{\rm gal}$ values, while the bulge masses are from \citealt{Ramsden25}]{Mummery_et_al_2024, MummeryVV25}, while the curve and shaded region is derived from first principles disk theory. 

The upper two panels show the disk-theory calibrated (but empirical) scaling relationships between the radiated energy/peak luminosity and the black hole mass. These curves are not based on a fundamental theory, but remain of practical use in constraining black hole masses from TDE data. It is clear to see visually that these scaling relationships describe the data extremely well, and therefore can be combined for a better method of constraining black hole masses.

\subsection{The {\tt TDEFLARE} model}
What remains is to construct a simple model which can be fit to data to extract $L_P, L_{\rm pk}$ and $E_g$ from a given TDE dataset, so that they can be combined in a statistically robust manner to constrain the black hole mass $M_\bullet$. The combining process is the subject of the next subsection, in this subsection we describe the simple model we fit to the data, which is based on that used in \cite{Mummery_et_al_2024}. 

We fit purely phenomenological models to multi wavelength TDE light curves. These phenomenological models are designed so as to most carefully extract the three parameters $L_P, L_{\rm pk}$ and $E_g$ of interest, and to include minimal additional nuisance parameters above what is required to constrain the parameters of interest. 

Specifically, these phenomenological models have the following functional form, there is a rise model
\begin{equation}
    L_{\rm rise}(\nu, t) = L_{\rm pk} \times f_{\rm rise}(t) \times \frac{\nu B(\nu, T)}{\nu_0 B(\nu_0, T)} , 
\end{equation}
and a decay model
\begin{equation}
    L_{\rm decay}(\nu, t) = L_{\rm pk} \times f_{\rm decay}(t) \times \frac{\nu B(\nu, T)}{\nu_0 B(\nu_0, T)} , 
\end{equation}
where $B(\nu, T)$ is the Planck function, and $\nu_0 = 6 \times 10^{14}$ Hz is a reference frequency (so that the $g$-band scaling relationships can be used directly). The early time SED of a TDE flare is almost certainly not described exactly by this blackbody profile, and this is simply included to make best use of multi wavelength data. The amplitude $L_{\rm pk}$ and temperature $T$ are common to both the rise and decay models. The functions $f_{\rm rise}(t)$ and $f_{\rm decay}(t)$ are defined so as to have a maximum amplitude of unity, so that $L_{\rm pk}$ remains the physical peak spectral luminosity amplitude. The rise model is that of a Gaussian rise
\begin{equation}\label{gauss_rise}
    f_{\rm rise}(t) = \exp\left( - {\left(t - t_{\rm peak}\right)^2 \over 2\sigma_{\rm rise}^2}\right) ,
\end{equation}
with fitting parameters $t_{\rm peak}$ and $\sigma_{\rm rise}$. The decay is modeled with an exponential 
\begin{equation}\label{pl_decay}
    f_{\rm decay}(t) = \exp\left(-{(t-t_{\rm peak})\over E_g/L_{\rm pk}}\right),
\end{equation}
with fitting parameters $t_{\rm peak}$ and $E_g$ (the radiated energy in the $g$-band exponential decay). The parameter $t_{\rm peak}$ is by default assumed to be common between the rise and decay models, and if no rise model is specified the parameter is fixed to $t_{\rm peak} = 0$.  The choice of the exponential decay is motivated by a desire to robustly measure the amplitude $L_{\rm pk}$, not to most accurately model the light curve shape themselves.  A (perhaps) more physically motivated power-law profiles  $(t-t_{\rm peak})^{-p}$  are poorly behaved in the  $t\to t_{\rm peak}$ limit, which often impacts the fitting process. 

It is important to stress at this point that the initial $\sim$ decade of decay in a given TDEs $g$-band luminosity really is well described by an exponential profile, as can be seen in Figure \ref{fig:standard}. The right hand panel shows that, when normalised by an amplitude and a simple exponential decay timescale, very different looking TDE light curves are all very similar over the first $\sim 3-5$ e-folding timescales. 

We also include a constant ``plateau'' luminosity at late ($t\geq t_{\rm peak}$) times, with 
\begin{equation}
    L_{\rm plat}(t) = L_P \times \frac{\nu B(\nu, T_P)}{\nu_0 B(\nu_0, T_P)} .
\end{equation}
The plateau luminosity $L_P$ is typically $\sim {\cal O}(1\%)$ of the peak luminosity $L_{\rm pk}$ \citep{Mummery_et_al_2024} so including this component at all times post-peak (rather than having it evolve naturally with time as a real accretion flow would)  induces minimal contamination into the values of the other parameters $L_{\rm pk}$ and $E_g$. 

The full light curve model is then 
\begin{multline}
    L(\nu, t) = L_{\rm rise}(\nu, t)\, \Theta(t_{\rm peak}-t) \\ + (L_{\rm decay}(\nu, t) + L_{\rm plat}(\nu, t))\, \Theta(t-t_{\rm peak}) ,
\end{multline}
where $\Theta(z) =1$ for $z\geq 0$, and $\Theta(z) = 0$ for $z < 0$. 

This profile is trivial to fit to data. We use the {\tt FitTeD} \citep{mummery2024fitted} package to perform the likelihood inference, which uses a chi-squared likelihood and the {\tt emcee} Monte Carlo Markov Chain code of \cite{EMCEE}.

\subsection{Black hole mass inference}
By fitting the above model to TDE data, we can constrain (up to) three parameters which depend strongly on black hole mass. What this in effect means is that we have (up to) three black hole mass distributions, $p(M_\bullet | L_{\rm pk})$, $p(M_\bullet | E_{\rm g})$ and $p(M_\bullet | L_{P})$ (where the notation $p(A|B)$ means the probability density of variable $A$ given measurement $B$), and we wish to answer the question ``what is the probability density function of $M_\bullet$ given this information?''.

In general, one could consider multiple approaches to answering this question. Common (but statistically incorrect) approaches include (i) averaging the data (for example, one might average 3 different velocity dispersion measurements of a TDE host and then use $M_\bullet-\sigma$ once), or (ii) averaging the posteriors (for example taking the mean of the $M_\bullet-\sigma$ and $M_\bullet-M_{\rm gal}$ values of the black hole mass). These two common approaches have a range of unfortunate statistical properties and are not in general recommended \citep{Hill11, Hill11a}. 

Instead, the recommended statistical approach is a process called ``conflation'', whereby one takes the product of the different probability density functions and renormalizes. This procedure can only be performed on {\it independent} statistical variables, of which we have two (the peak luminosity and radiated energy are not independent). Explicitly then our mass inference is 
%\begin{multline}
%    p(M_\bullet| L_{\rm pk}, E_g, L_P) \\ = {p(M_\bullet | L_{\rm pk})\, p(M_\bullet | E_{\rm g})\, p(M_\bullet | L_{P}) \over \int p(M_\bullet | L_{\rm pk}) \, p(M_\bullet | E_{\rm g})\, p(M_\bullet | L_{P})\, {\rm d}M_\bullet  }. 
% \end{multline}
\begin{equation}
p(M_\bullet| E_g, L_P)  = {p(M_\bullet | E_{\rm g})\, p(M_\bullet | L_{P}) \over \int  p(M_\bullet | E_{\rm g})\, p(M_\bullet | L_{P})\, {\rm d}M_\bullet  }. 
\end{equation}
We make the choice of the radiated energy here as it has smaller intrinsic scatter, this however is personal preference and the peak luminosity scaling could be used. 
This method has a number of favorable statistical properties, namely it is the unique procedure which (i) minimizes the loss of Shannon Information when combining the probability density functions, (ii) it is the best linear unbiased estimator of the true distribution, and (iii) is the maximum likelihood estimator of the true distribution \citep{Hill11, Hill11a}.  

One of the important properties of conflation (relevant for the problem at hand), is that the conflation of $N$ Gaussian's remains a Gaussian, with a mean $\mu$ and variance $\sigma^2$ given by \citep{Hill11}
\begin{align}
    \mu &= \sum_{i=1}^N {\mu_i/\sigma_i^2 \over 1/\sigma_i^2}, \\
    \sigma^2 &= \sum_{i=1}^N {1 \over 1/\sigma_i^2} .
\end{align}
This is simple to prove for $N=2$ (it is just completing the square within the exponential) and the generalization to $N>2$ follows from performing the same factorization term by term. 

Each of our probability density functions are log-normal distributions (i.e., they are normal distributions on a log scale), or specifically for each ``observable'' ${\cal O} \in \{L_{\rm pk}, E_g, L_P\}$ we have 
\begin{multline}
    p(\log M_\bullet|{\cal O})\,= \\  {1\over \sqrt{2\pi s_{\cal O}^2}} \exp\left(-{(\log \mu_{M_\bullet}({\cal O})-\log M_\bullet)^2\over 2s_{\cal O}^2}\right)\, .
\end{multline}
Here $\log \mu_{M_\bullet}({\cal O})$ is the (logarithm of the) black hole mass given by the scaling relationship (for inferred light curve parameter ${\cal O}$), and 
\begin{equation}
    s_{\cal O}^2 = \epsilon_{\cal O}^2 + \sigma_{\cal O}^2 ,
\end{equation}
where we combine the intrinsic scatter in each scaling relationship $\epsilon_{\cal O}$ in quadrature with the variance in the inferred parameter from the  fit to the data $\sigma_{\cal O}^2$. 

What this means practically is that our conflated distribution will remain log-normal but with a scatter that is {\it reduced} in log-space (this is intuitively obvious, the more statistical information one has about a variable the more precise one can be in its value). Let us imagine that all observables were measured perfectly from the data (i.e., $\sigma_{\cal O}^2=0$), then our black hole mass uncertainty would become (in log space)
\begin{equation}
    \sigma = \sqrt{{1\over %\epsilon_{L_{\rm pk}}^{-2} + 
    \epsilon_{E_g}^{-2} + \epsilon_{L_P}^{-2}}} \approx  0.28\, {\rm dex}, 
\end{equation}
for %$\epsilon_{L_{\rm pk}}=0.53\, {\rm dex}$, 
$\epsilon_{E_g} = 0.44\, {\rm dex}$ and $\epsilon_{L_P} = 0.38\, {\rm dex}$ (the values assumed in {\tt TDEFLARE}). Of course, measurement uncertainty on the various light curve parameters will increase the uncertainty here, but this highlights the point that by combining scaling relationships one can make increasingly precise inference on TDE black hole masses. 

The above approach however neglects a key piece of relativistic TDE physics, of importance at the high black hole mass end of the TDE population. This is the presence of the so-called \cite{Hills75} mass, which represents the black hole mass at which a given star with mass $M_\star$ and radius $R_\star$ would be disrupted precisely at the event horizon of a black hole with spin $a_\bullet$, having come in on an orbit with a velocity vector which made an asymptotic angle $\psi$ with the equatorial plane (tidal forces depend on inclination in general relativity). The fully relativistic (Kerr metric) Hills mass was derived in \cite{Mummery24}, and is equal to 
\begin{multline}\label{HillsMass}
    M_{\rm Hills}(a_\bullet, \psi, M_\star, R_\star) = \left[{2 R_\star^3 c^6 \over \eta G^3 M_\star } \right]^{1/2} {1\over \chi^{3/2}} \\ 
     \Bigg[1 + {6\chi \over \chi^2 - a_\bullet^2 \cos^2\psi} + {3a_\bullet^2 \over 2\chi^2} - {6a_\bullet \sin \psi \over \sqrt{\chi^3 - a_\bullet^2 \chi \cos^2\psi}} \Bigg]^{1/2}, 
\end{multline}
where $\chi(a_\bullet, \psi)$ is the location of the innermost bound spherical orbit (normalised by $GM_\bullet/c^2$), which is given by the solution of the following octic polynomial \citep{Hod13} 
\begin{multline} \label{chi_eq_1}
    \chi^4 - 4\chi^3 - a_\bullet^2(1 - 3 \cos^2\psi)\chi^2 + a_\bullet^4\cos^2\psi \\ + 4a_\bullet \sin \psi \sqrt{\chi^5 - a_\bullet^2\chi^3\cos^2\psi} = 0 .
\end{multline}
The dimensional prefactor $\left({2 R_\star^3 c^6 /  G^3 M_\star }\right)^{1/2} \approx 2.5 \times 10^8 M_\odot$, which highlights the broad scale at which relativistic Hills mass effects become important. Finally, $\eta$ is a dimensionless nuisance factor relating to the strength of the self-gravitational acceleration of the star (in units of $GM_\star/R_\star^2$). %We take $\eta = 1$ in this work. 

Incorporating the effects of the Hills suppression into our black hole mass inference is simple enough, one simply adopts a Bayesian framework and marginalizes over all possibly stellar masses, spins and incoming inclination angles, only keeping those systems with black hole masses below the Hills mass (i.e., those disruptions that can produce detectable emission). To be explicit, one computes the final black hole mass probability density function from 
\begin{multline}
    p(M_\bullet) = \iiint p(M_\bullet|%L_{\rm pk}, 
    E_g, L_P)\, {\cal H}(M_\bullet, M_\star, a_\bullet, \psi) \\ p(M_\star) p(a_\bullet) p(\psi) \, {\rm d}M_\star\, {\rm d}a_\bullet \, {\rm d}\psi . 
\end{multline}
The Hills factor ${\cal H}$ in this integral is simply equal to $1$ if $M_\bullet\leq M_{\rm Hills}$ and zero otherwise. One must assume a mass-radius relationship to compute this integral, and priors on the other parameters $(p(M_\star), p(a_\bullet), p(\psi))$. The publicly available code {\tt tidalspin} \citep{Mummery24} computes the above integral under the assumption that the black hole spins are uniformly distributed, that the stars in the galactic center are isotropically distributed and have a mass distribution of a \cite{Kroupa01} initial mass function multiplied by the \cite{Wang04} TDE rate as a function of stellar properties. The stars are assumed to follow the \cite{Kippenhahn90} mass-radius relation. We use identical assumptions here. 

By marginalizing over the physics of the Hills mass, the output probability density function will no longer be log-normal (as the above integral does not preserve normality). However, unless the black hole mass is close to $M_{\rm Hills}$, the posterior distribution remains very well approximated by a log-normal.

\section{Implementation, examples and results}\label{implementation}
\subsection{A simple example: AT2019dsg}
We begin with an explicit example, fitting the optical/UV  data of the TDE AT2019dsg. We are motivated to use AT2019dsg as a first example as it (i) has excellent multi-wavelength data, (ii) has independent estimates of the black hole mass from galactic scaling relationships, and (iii) has been modeled with the full {\tt FitTeD} disk code  \citep{mummery2024fitted} and the {\tt kerrSED} disk spectral fitting code \citep{GuoloMum24, Guolo25c}, which will both act as interesting points of comparison. 

\begin{figure*}
    \centering
    \includegraphics[width=0.53\linewidth]{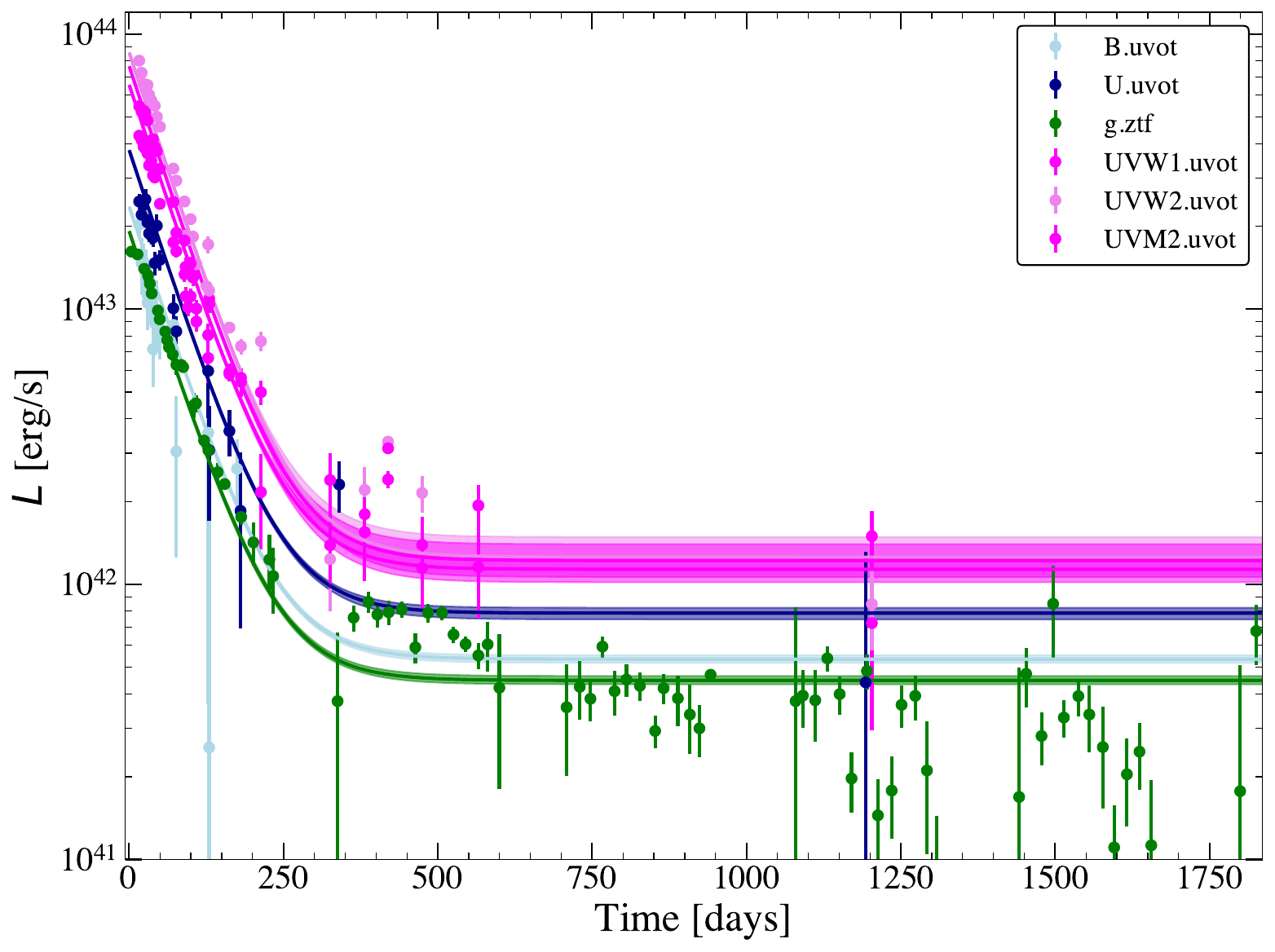}
    \includegraphics[width=0.4\linewidth]{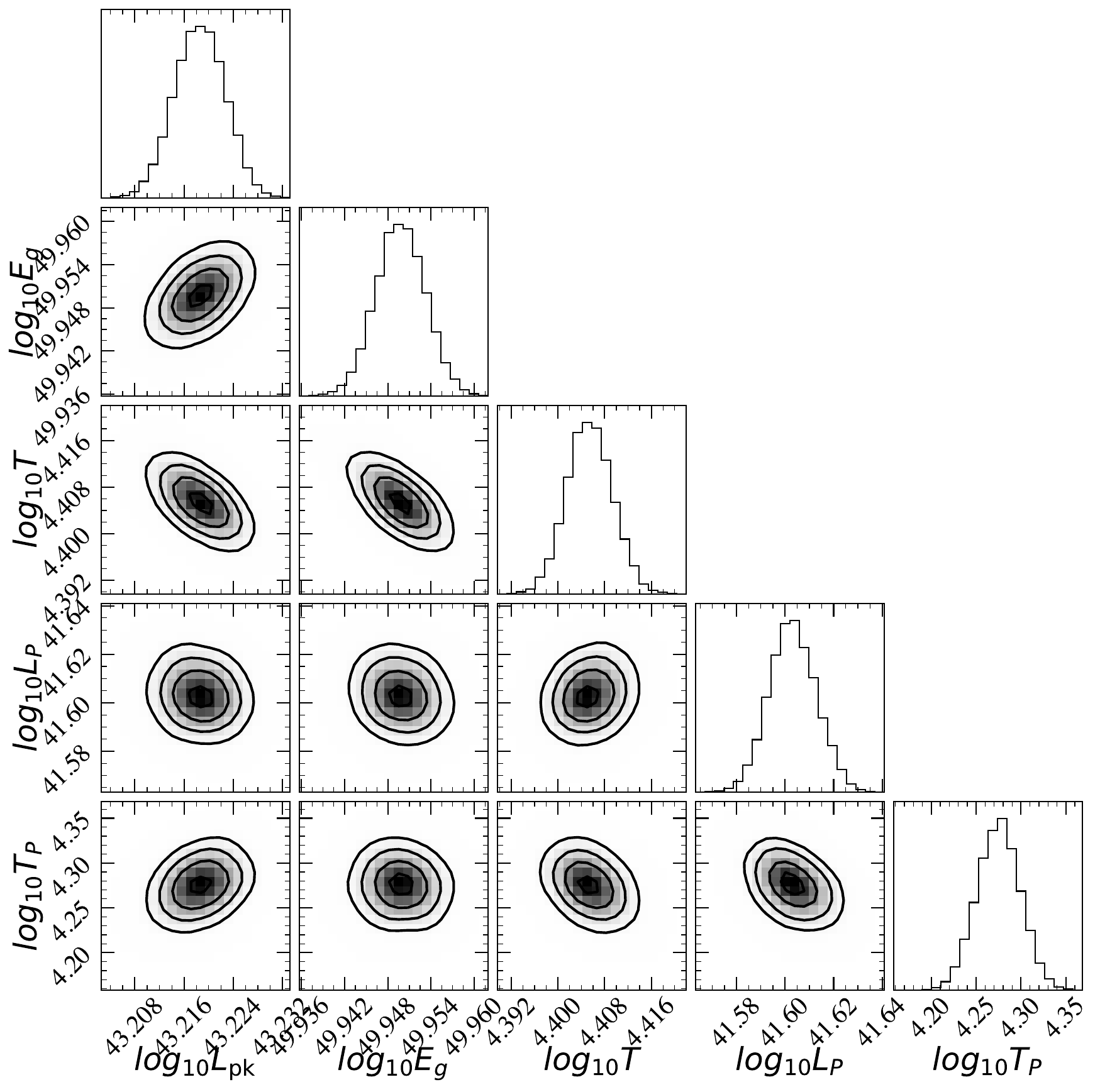}
    \includegraphics[width=0.45\linewidth]{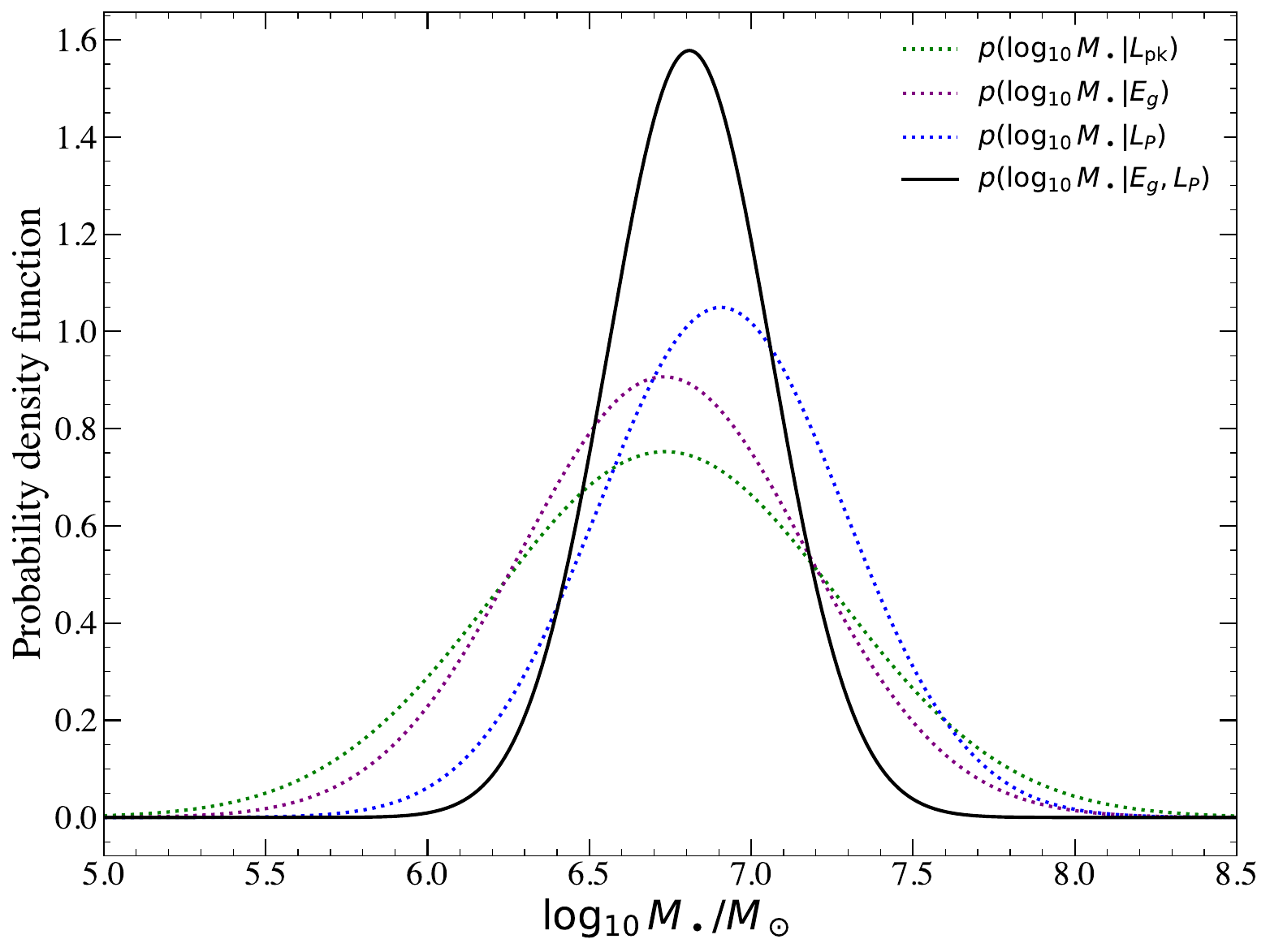}
    \includegraphics[width=0.45\linewidth]{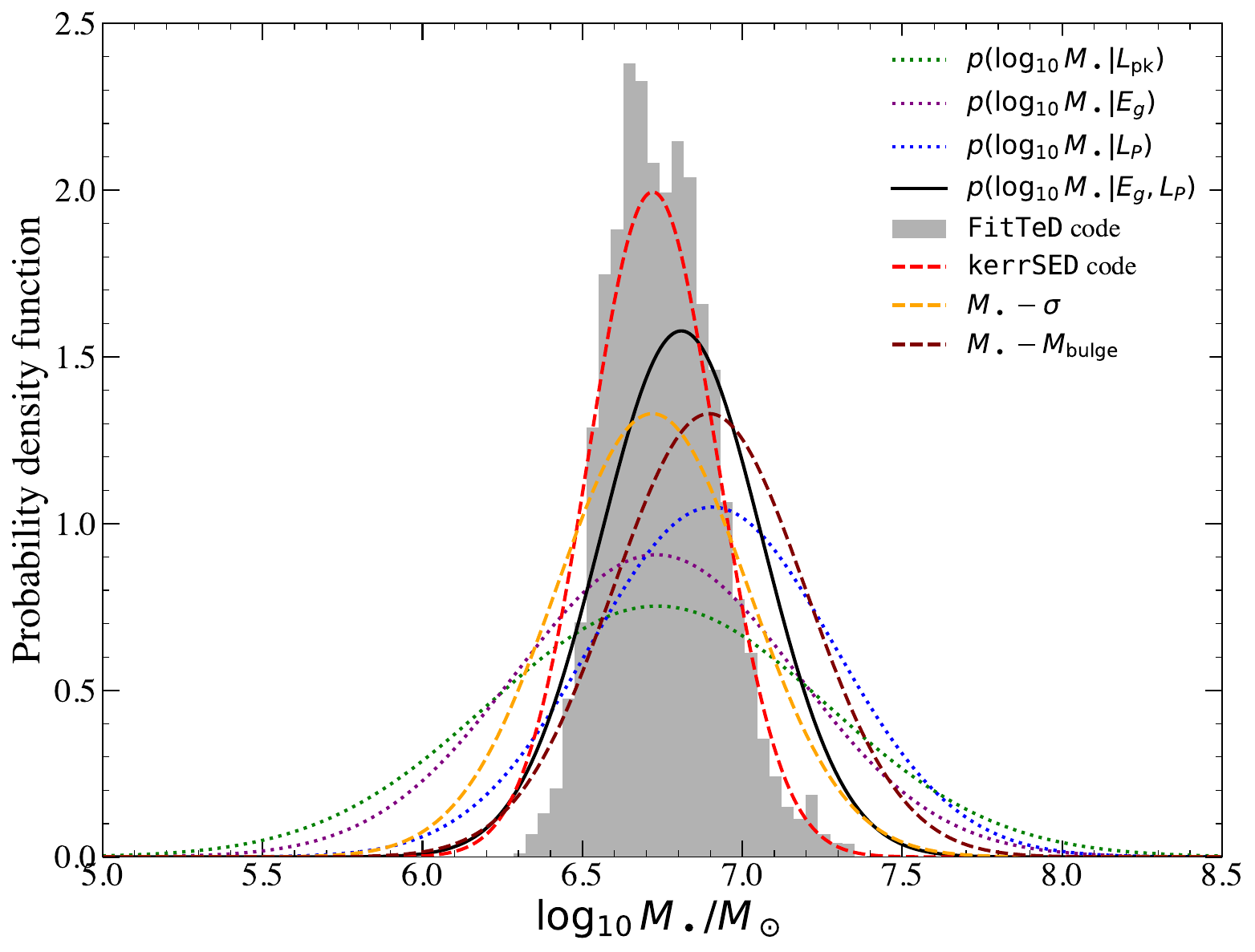}
    \caption{An example of the {\tt TDEFLARE} fitting process and results. Upper left: the observed and modeled multi wavelength light curves of AT2019dsg, highlighting that the phenomenological model reproduces the data well. Upper right: the posterior distributions of the fitted parameters, showing that the three important parameters can be constrained well. Lower left: the three individual posterior distributions of the black hole mass, from
    the different observed parameters (blue, green and purple) and the conflated distribution (black). Lower right: a comparison of various black hole mass inference techniques, with varying degrees of physical content. All are consistent at $1\sigma$. }
    \label{fig:19dsg}
\end{figure*}

In Figure \ref{fig:19dsg} we display the fit of the {\tt TDEFLARE} model to the multi wavelength data of AT2019dsg (upper left), and the corner plot of the resulting posterior distributions of the fitted parameters (upper right). The data is taken from the {\tt manyTDE} database\footnote{https://github.com/sjoertvv/manyTDE}. Clearly, the phenomenological model can reproduce the data and provide tight constraints on the mass scaling parameters $L_{\rm pk}, E_g$ and $L_P$. 

The individual black hole mass posteriors corresponding to these inferred parameters are shown by the blue, green and purple dotted curves in the lower left panel. The conflation of the energy and plateau posteriors is shown by the black solid curve (highlighting the increased precision which comes from the conflation process). We do not show the posterior after Hills mass impacts are taken into account, as it is indistinguishable from this black curve (this is simply because the inferred masses are low compared to the canonical Hills scale of $\sim 10^8 M_\odot$). 

We contrast these black hole mass estimates with approaches that contain significantly more physics ({\tt FitTeD} and {\tt kerrSED}) and galactic scaling relationships ($M_\bullet-\sigma$ and $M_\bullet-M_{\rm bulge}$) in the lower right panel. All are effectively identical, and are certainly indistinguishable. 

This is important result, as in addition to performing a much more detailed analysis of accretion disk emission, both {\tt FitTeD} and {\tt kerrSED} include X-ray emission in their parameter inference, either in a light curve fitting sense ({\tt FitTeD}) or a spectral fitting sense ({\tt kerrSED}). Clearly this X-ray emission is not essential to constraining the {\it mass} of the black hole, a result of the mass inference (in all approaches) being driven by the amplitude of the plateau flux. This is certainly a convenient result as we enter the LSST era. 

We wish to stress that when X-ray data is available, it should be used in constraining the parameters of the TDE system. The purpose of this example was to show that {\tt TDEFLARE} produces results which are comparable to much more detailed physical models, {\it it is not the suggestion of the author that it should be used instead of these more physical models when it is possible to do so.}

\begin{figure*}
    \centering
    \includegraphics[width=0.53\linewidth]{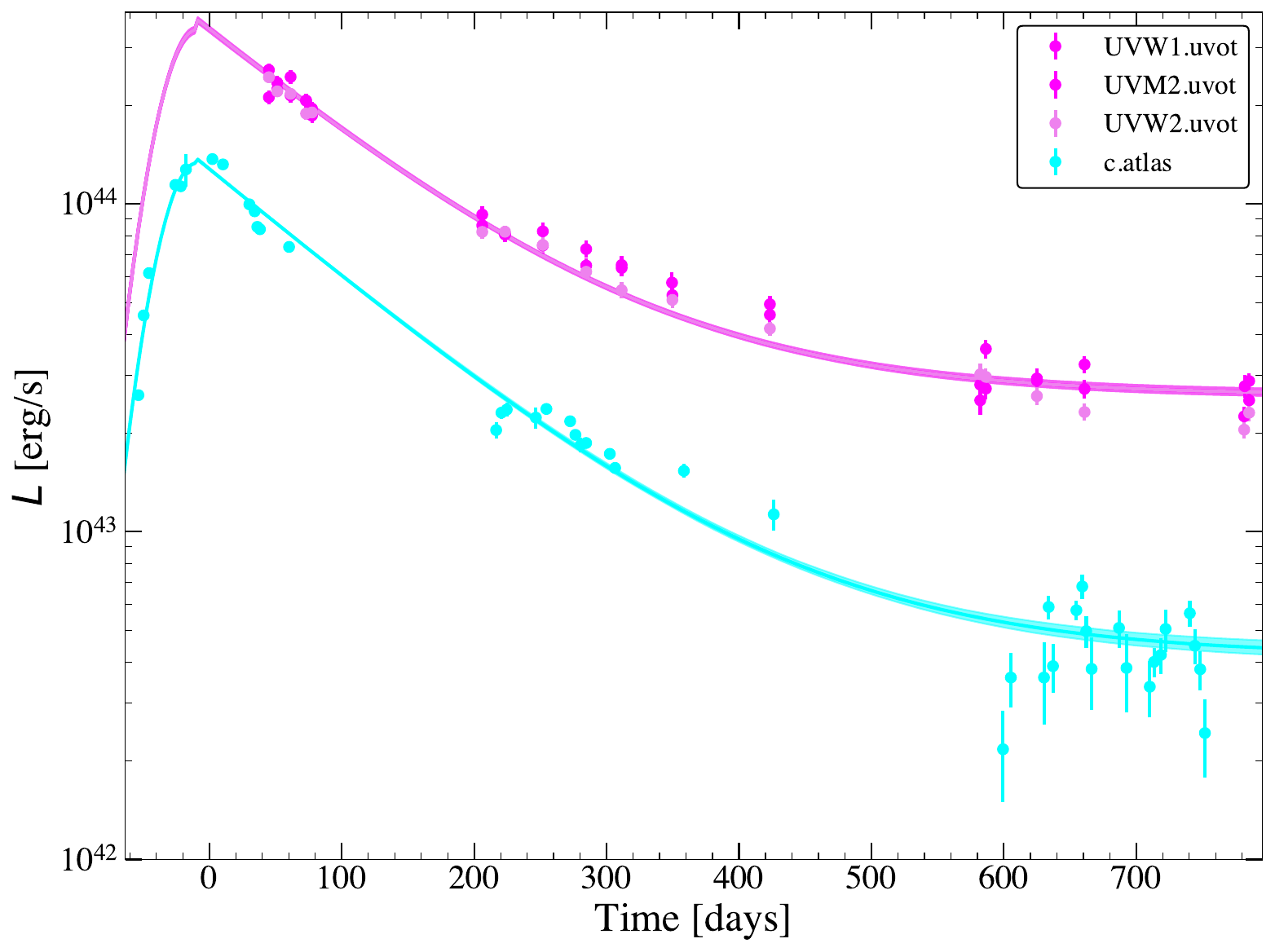}
    \includegraphics[width=0.4\linewidth]{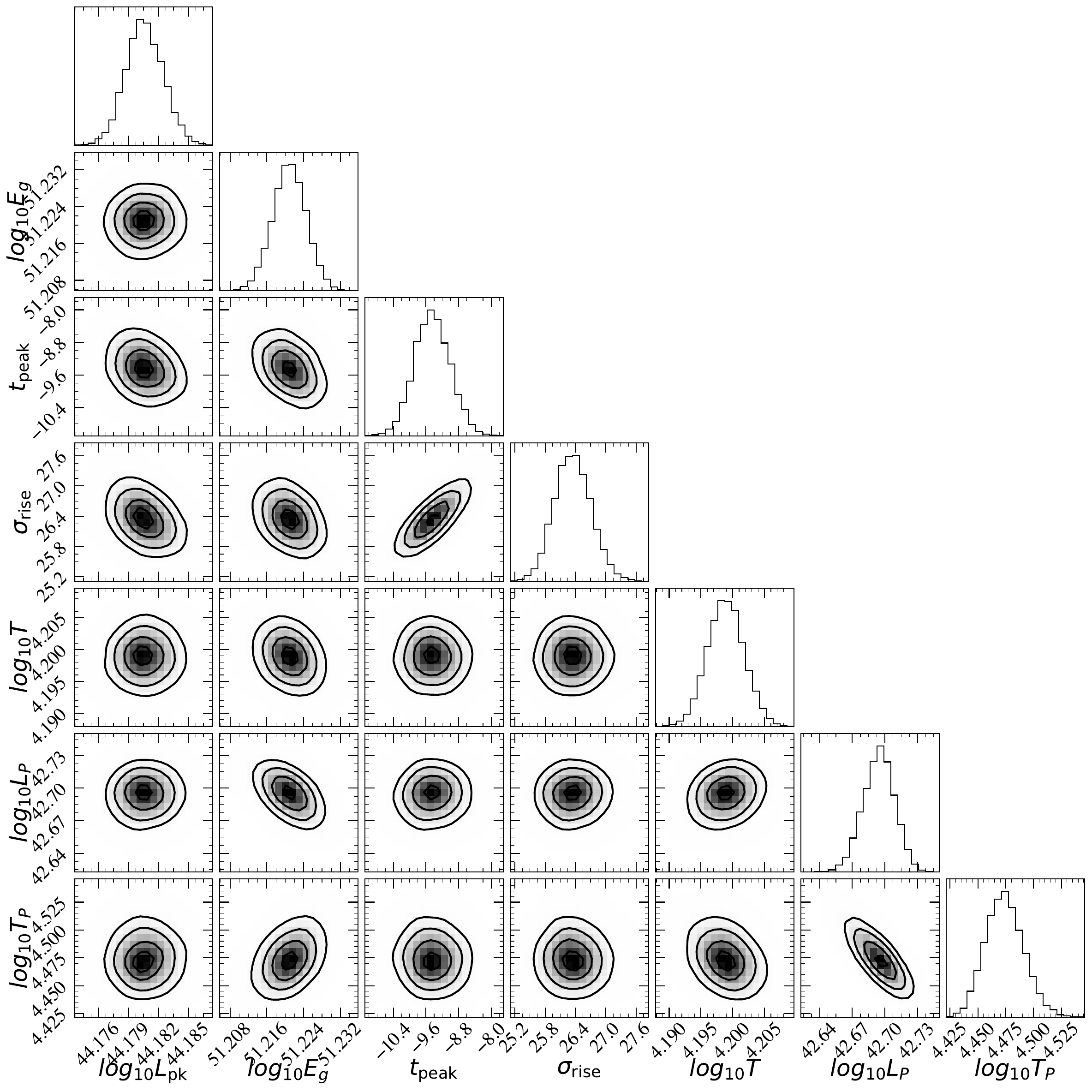}
    \includegraphics[width=0.65\linewidth]{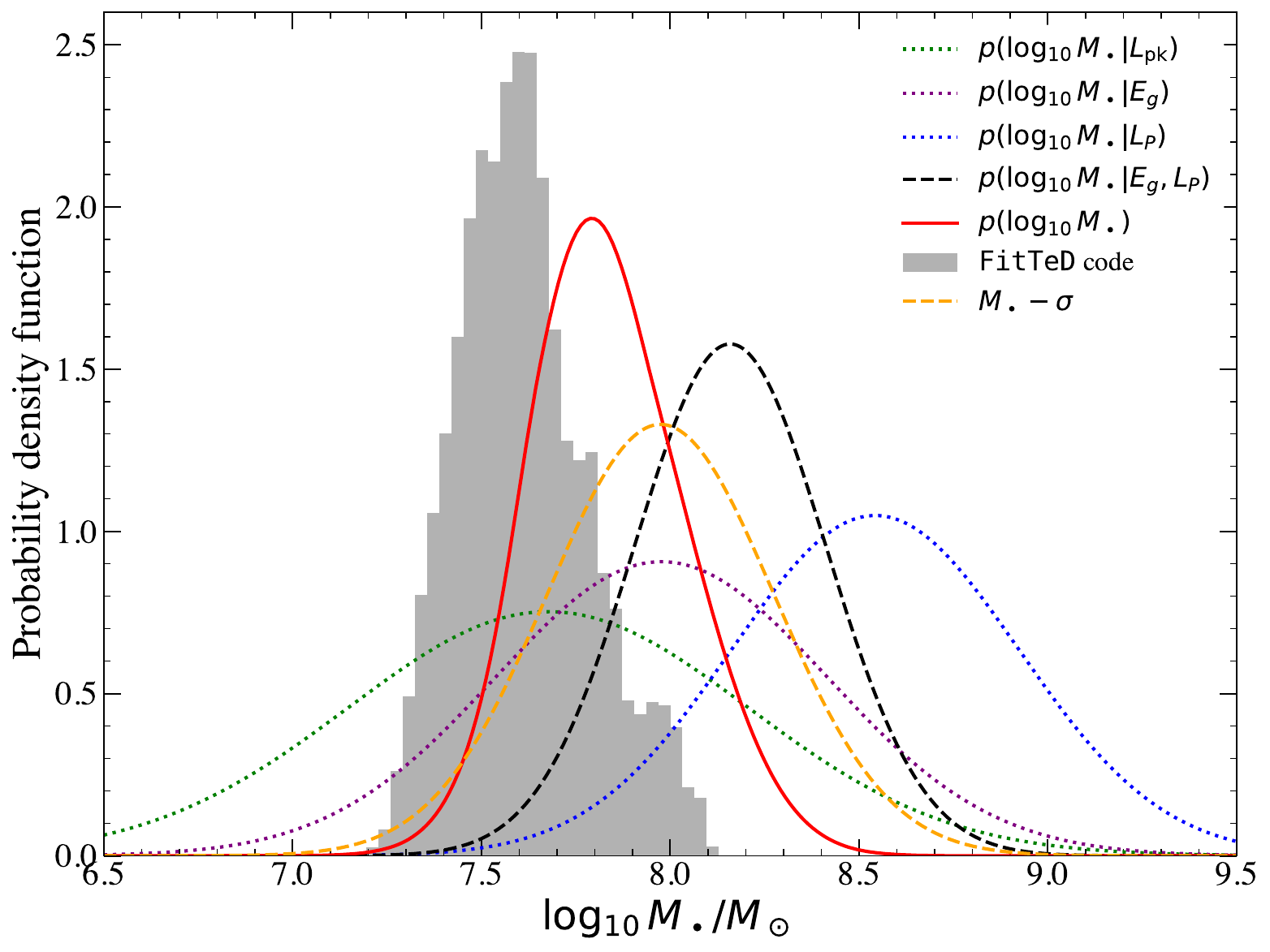}
    \caption{The importance of including the physics of the Hills mass when constraining the black hole mass of bright (high mass) sources. The upper panel shows the light curve fits and posterior distributions of fits to the TDE eJ2344. The source is extremely bright, implying high mass posteriors (lower panel). If one does not include Hills mass physics one infers a mass which is above the typical Hills scale (black curve), when Hills mass physics is taken into account the mass posterior drops (red curve), bringing it into line with more  complex models (numerical posterior from the {\tt FitTeD} code shown as a grey histogram). The posterior is also consistent with the $M_\bullet-\sigma$ relationship (orange dashed curve).    }
    \label{fig:eJ2344}
\end{figure*}
\begin{figure*}
    \centering
    \includegraphics[width=0.4\linewidth]{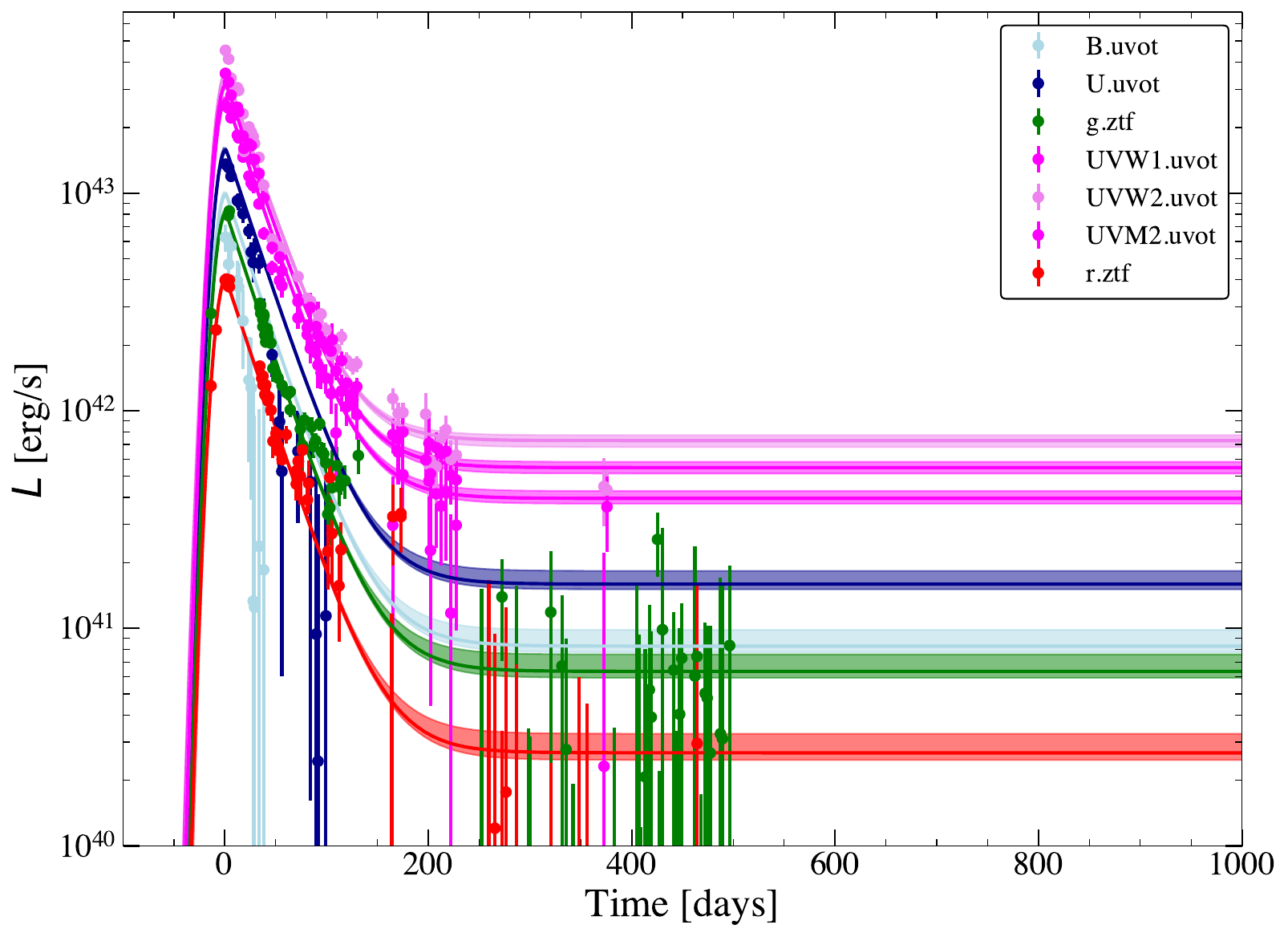}
    \includegraphics[width=0.4\linewidth]{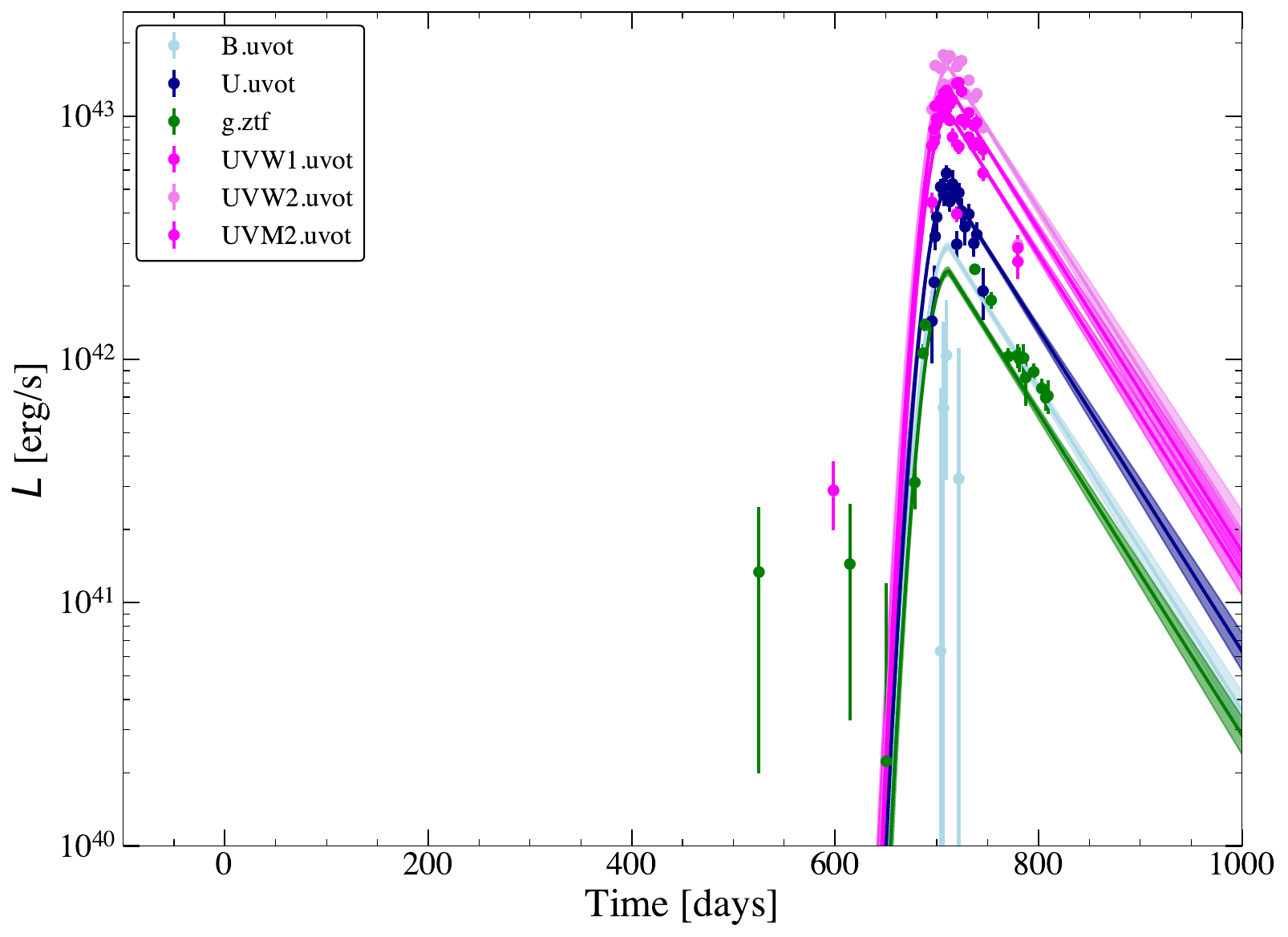}
    \includegraphics[width=0.35\linewidth]{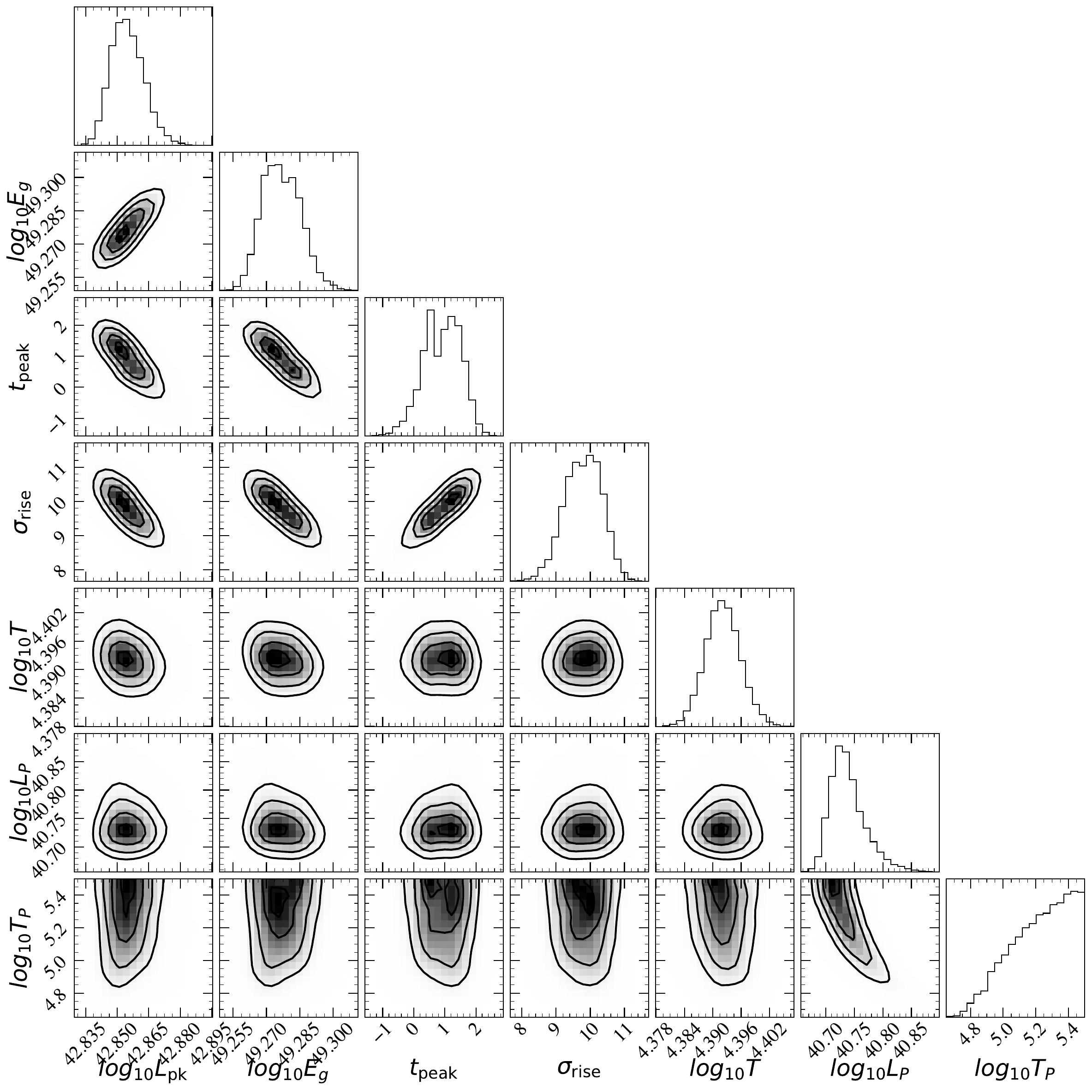}
    \includegraphics[width=0.35\linewidth]{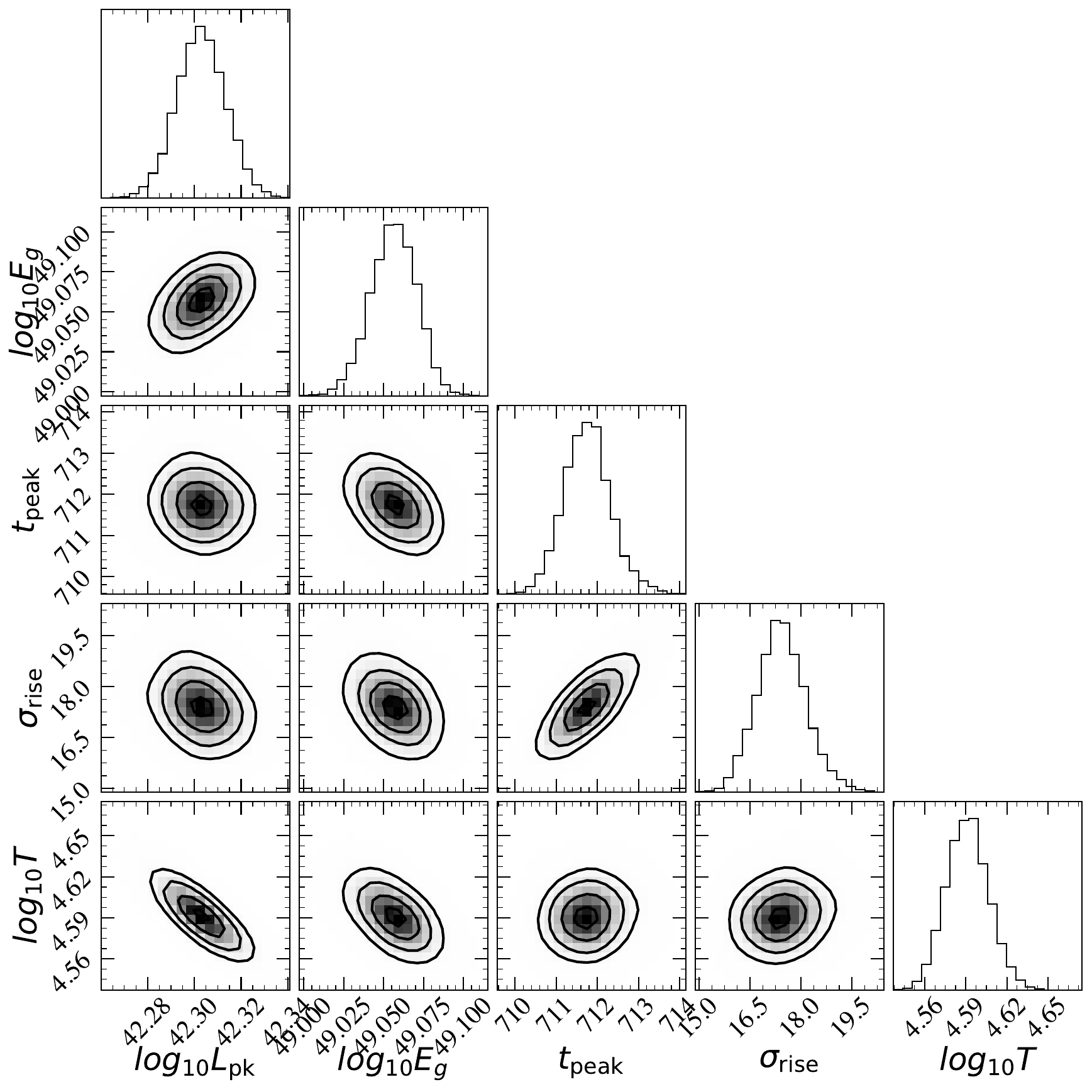}
    \includegraphics[width=0.5\linewidth]{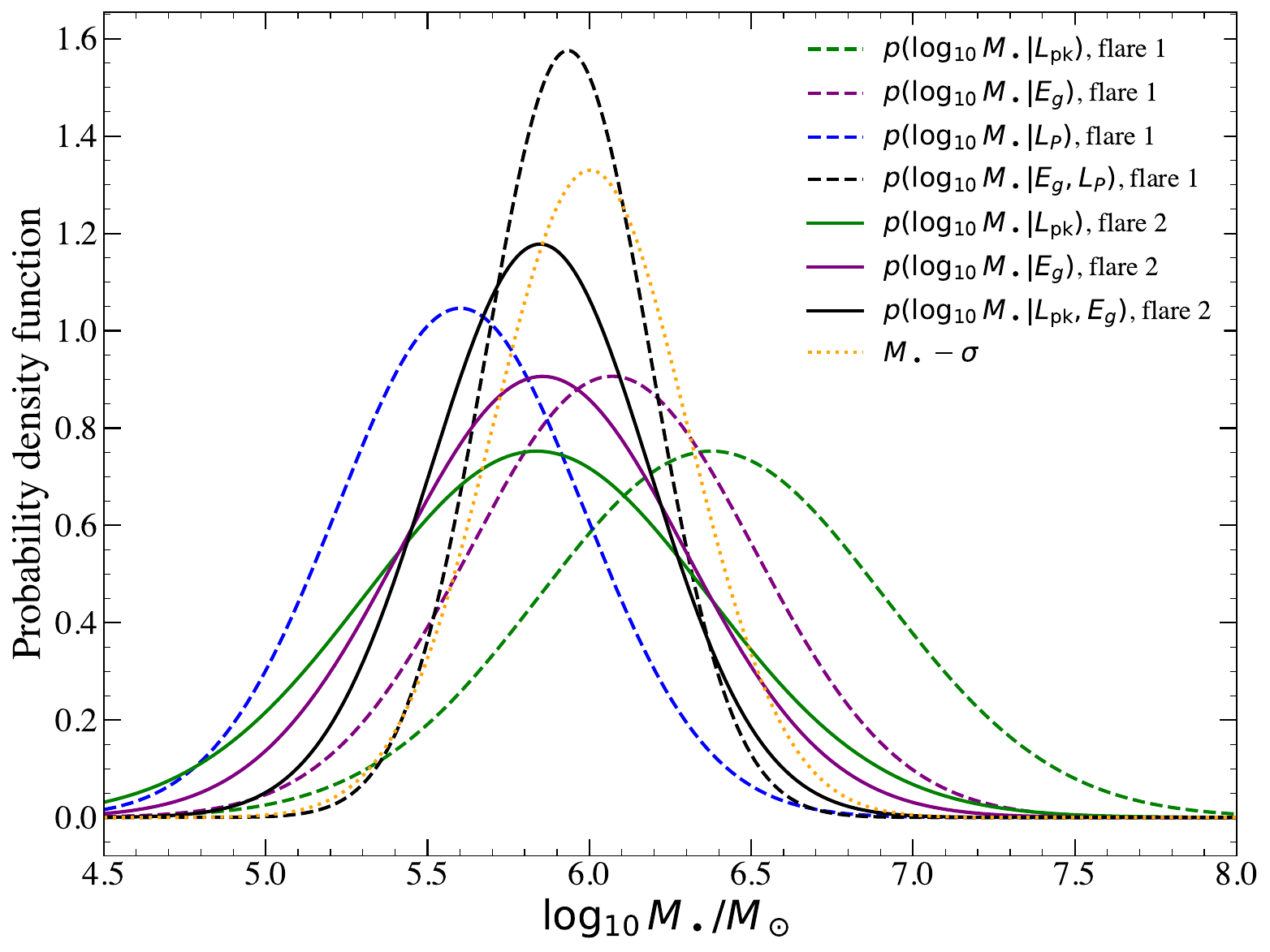}
    \caption{Fitting {\tt TDEFLARE} to two successive flares of the repeating partial tidal disruption event AT2022dbl. The upper left and right panels show light curve modeling of the two successive flares, with parameter posteriors shown in the middle row. The first flare has sufficient data to constrain the plateau, while the second does not. The lowest panel shows black hole mass inference from each scaling relationship (green, purple and blue) techniques for flare 1 (dashed curves) and flare 2 (solid curves). The conflated distributions (black) are consistent between the two epochs, and are consistent with the $M_\bullet-\sigma$ mass at $1\sigma$.  }
    \label{fig:22dbl}
\end{figure*}

\subsection{A high mass example: eJ2344}
The previous example showed that the premise of the model worked, but did not highlight the importance of the Hills mass in constraining TDE properties at the high mass end. To show this, we turn to the high mass TDE eRASSt J234402.9-352640 \citep{Homan2023_eJ2344}, which we shall henceforth refer to as ``eJ2344''. All multi-wavelength data is taken from \cite{Homan2023_eJ2344}. This TDE was chosen as it is one of the small number of TDEs which are high mass and has bright X-ray emission, and so can be (and has been) modeled with {\tt FitTeD}, allowing a detailed comparison.

In Figure \ref{fig:eJ2344} we again show the light curve fit (upper left), and corner plot (upper right). Note that by virtue of being a high mass TDE, eJ2344 is extremely bright (it is also very bright in the X-rays, with $L_X \sim 10^{44}$ erg/s at peak, \citealt{Homan2023_eJ2344}). This means that the black hole mass posteriors are high (bottom panel). 

The lower panel highlights the importance of the Hills mass -- in red we display the mass posterior once the event horizon suppression is accounted for (the black curve neglects this physics). We see that including Hills suppression shifts the mass posterior to lower values (and better in keeping with the {\tt FitTeD} posterior, which includes all relativistic physics and models the X-ray emission). By virtue of being a fully relativistic disk model, {\tt FitTeD} takes into account the Hills mass physics by default, while {\tt TDEFLARE} has to input it ``by hand'' (again highlighting the benefit of fitting more physically detailed models when possible). As is the case with AT2019dsg, all of these mass inference techniques are consistent with the $M_\bullet-\sigma$ value. 

\subsection{Repeated flares in partial TDEs: AT2022dbl}
Not every TDE will be a ``full'' (complete) disruption and many will instead be partial disruptions, with a stellar remnant surviving. For a subset of partial TDEs this remnant will return and be disrupted again, producing a second (or more) flare(s).  It is important to check that {\tt TDEFLARE} returns sensible values for the black hole mass in these cases. Of course, one can only check that (i) the mass inference is consistent between flares (as the black hole mass certainly will not change), and (ii) the mass inference is consistent with some galactic scaling relationship (like $M_\bullet-\sigma$).

\begin{figure*}
    \centering
    \includegraphics[width=0.35\linewidth]{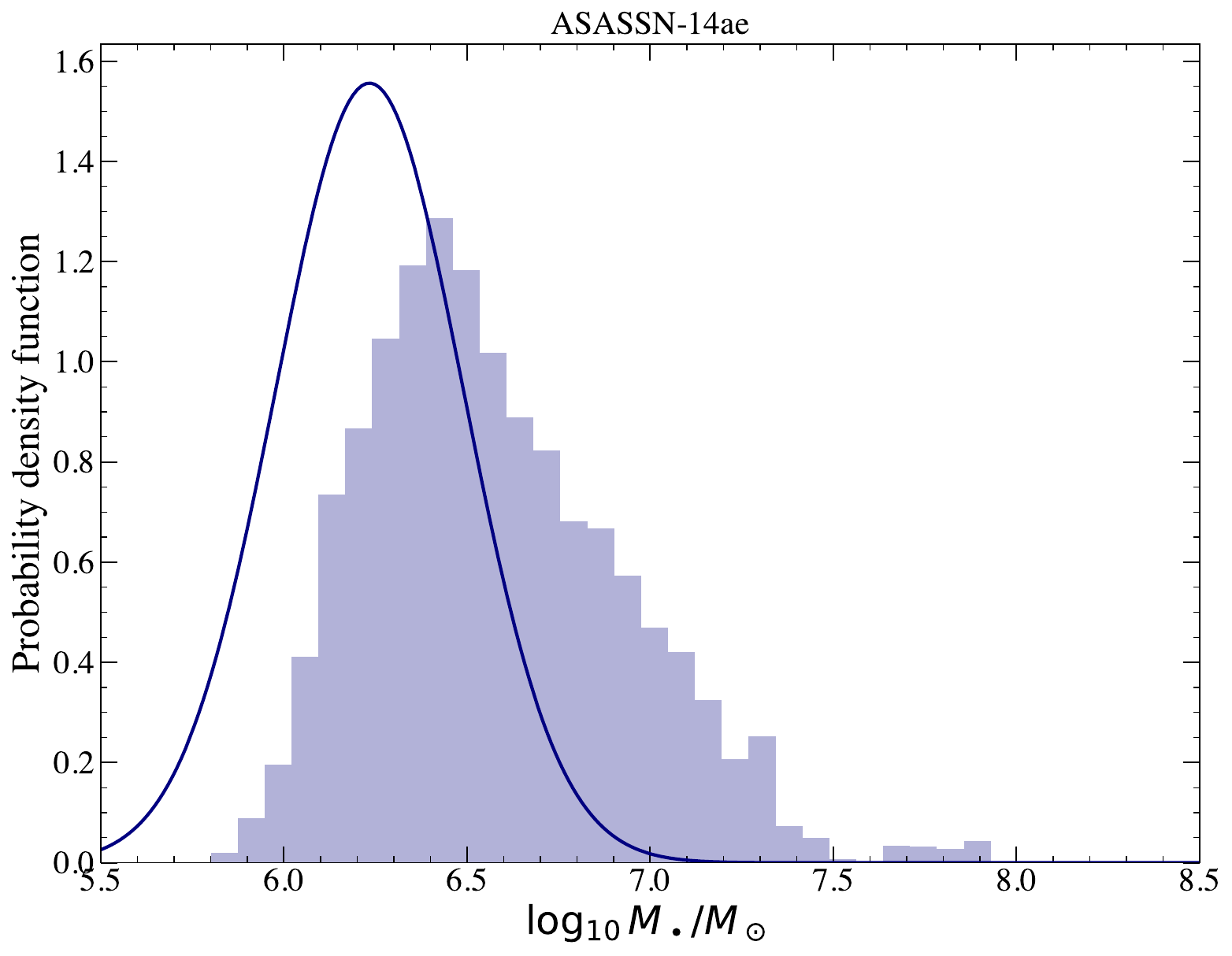}
    \includegraphics[width=0.35\linewidth]{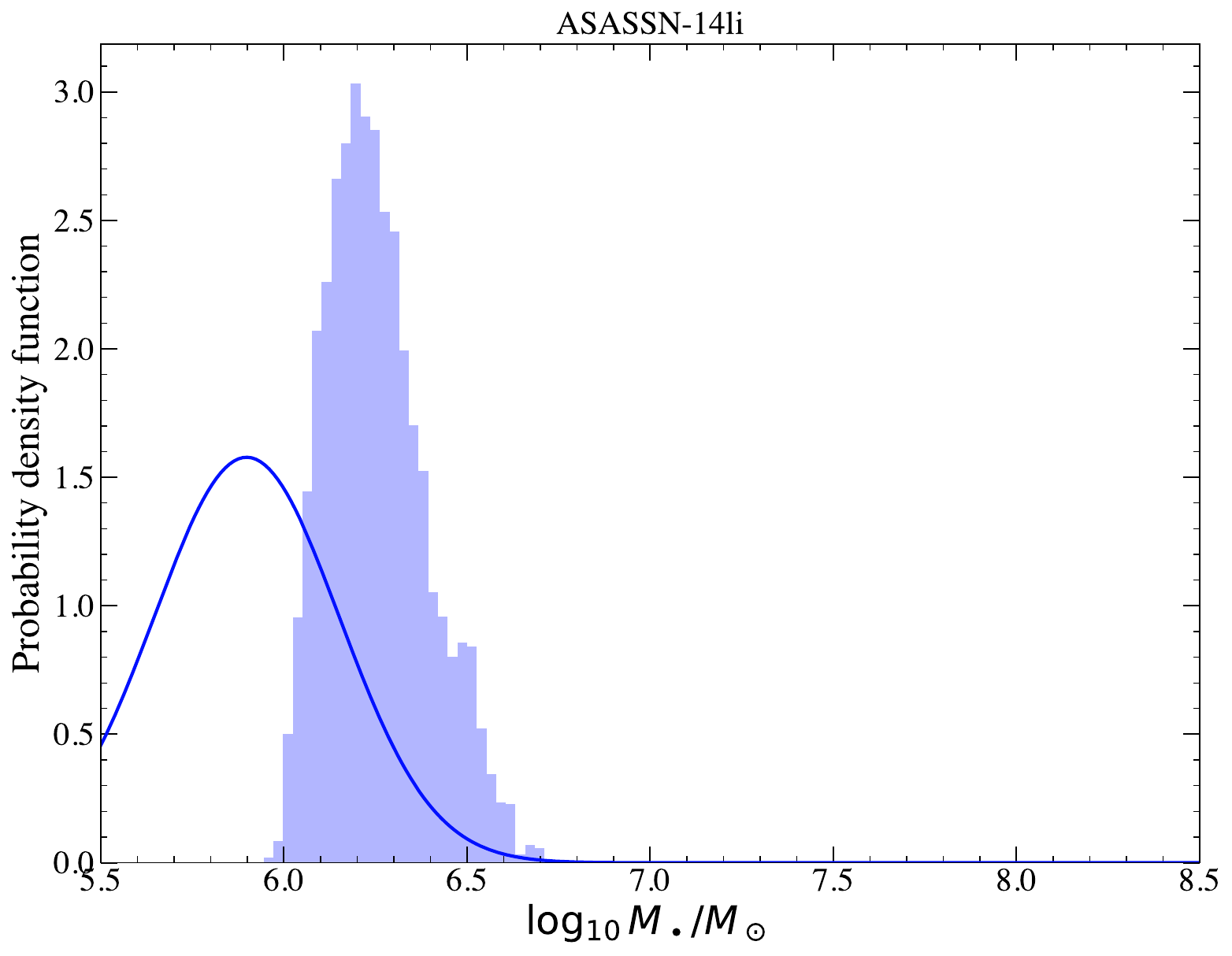}
    \includegraphics[width=0.35\linewidth]{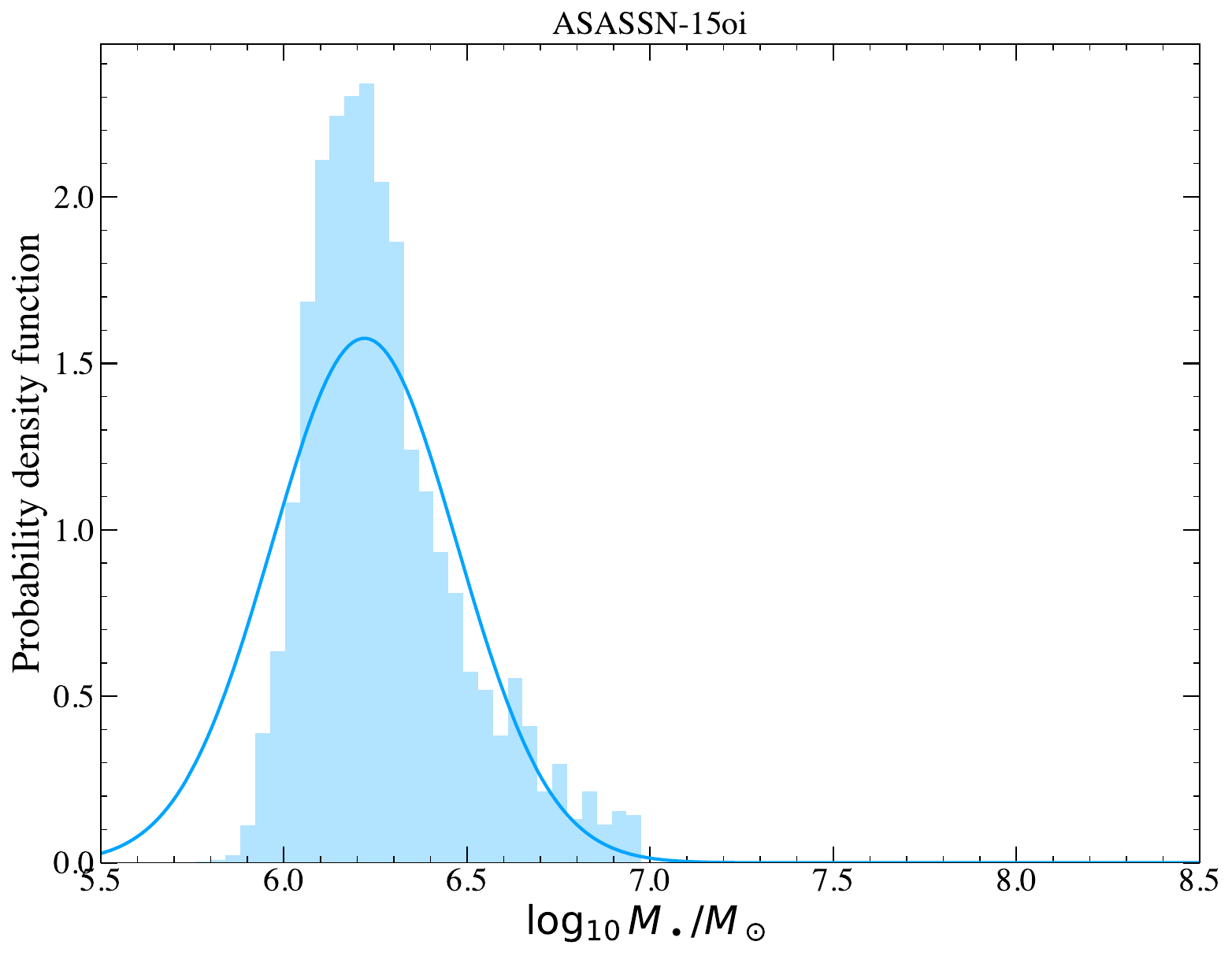}
    \includegraphics[width=0.35\linewidth]{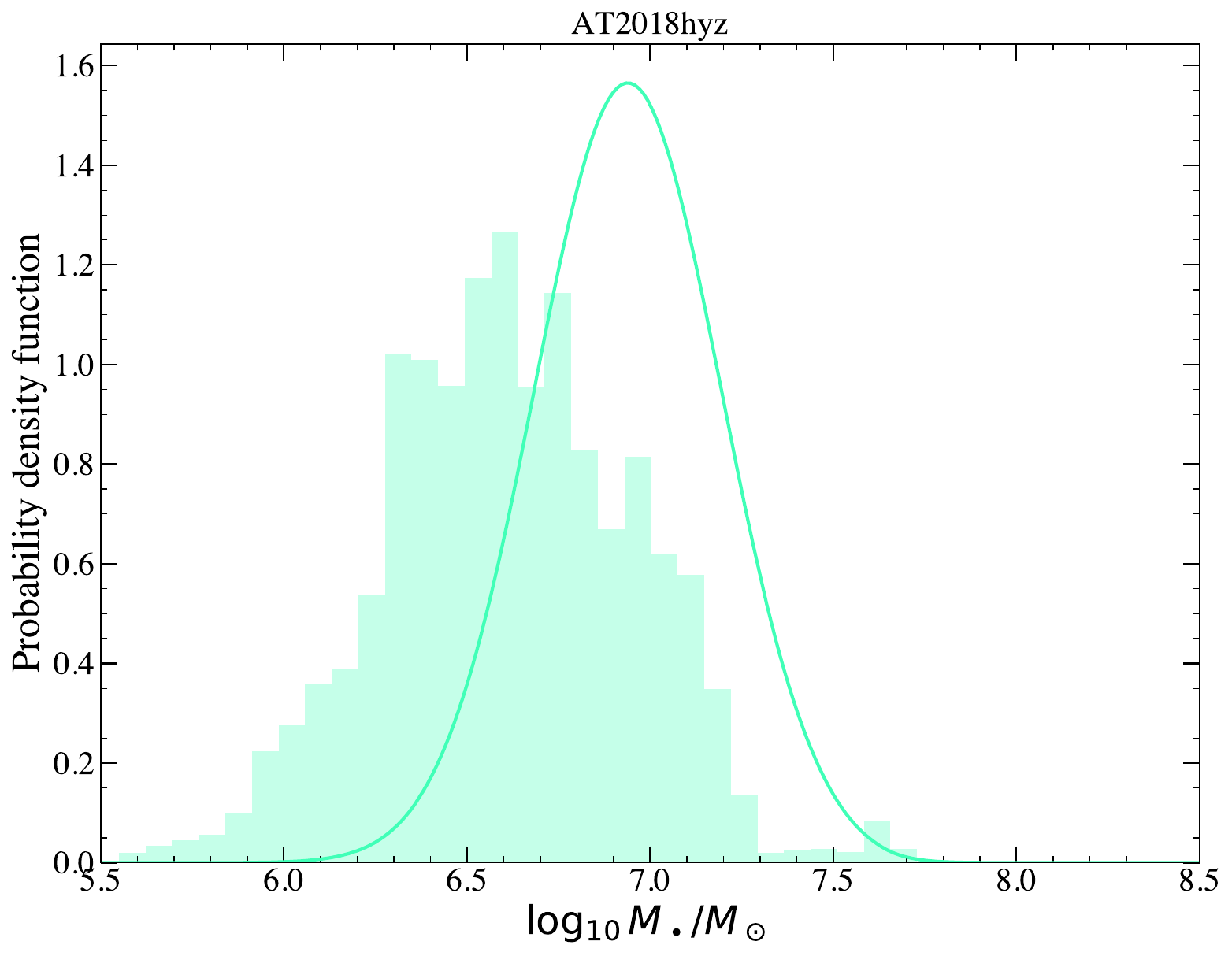}
    \includegraphics[width=0.35\linewidth]{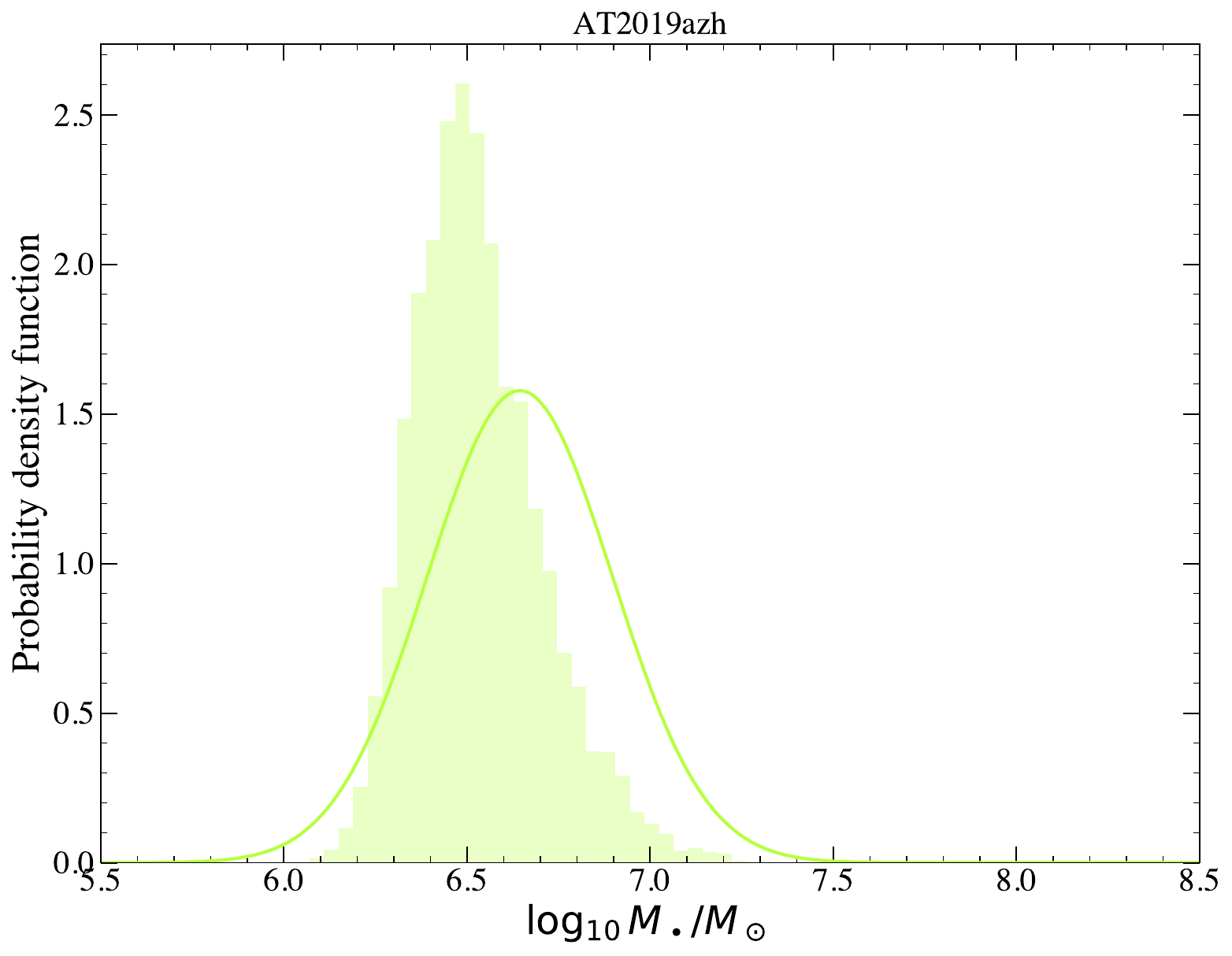}
    \includegraphics[width=0.35\linewidth]{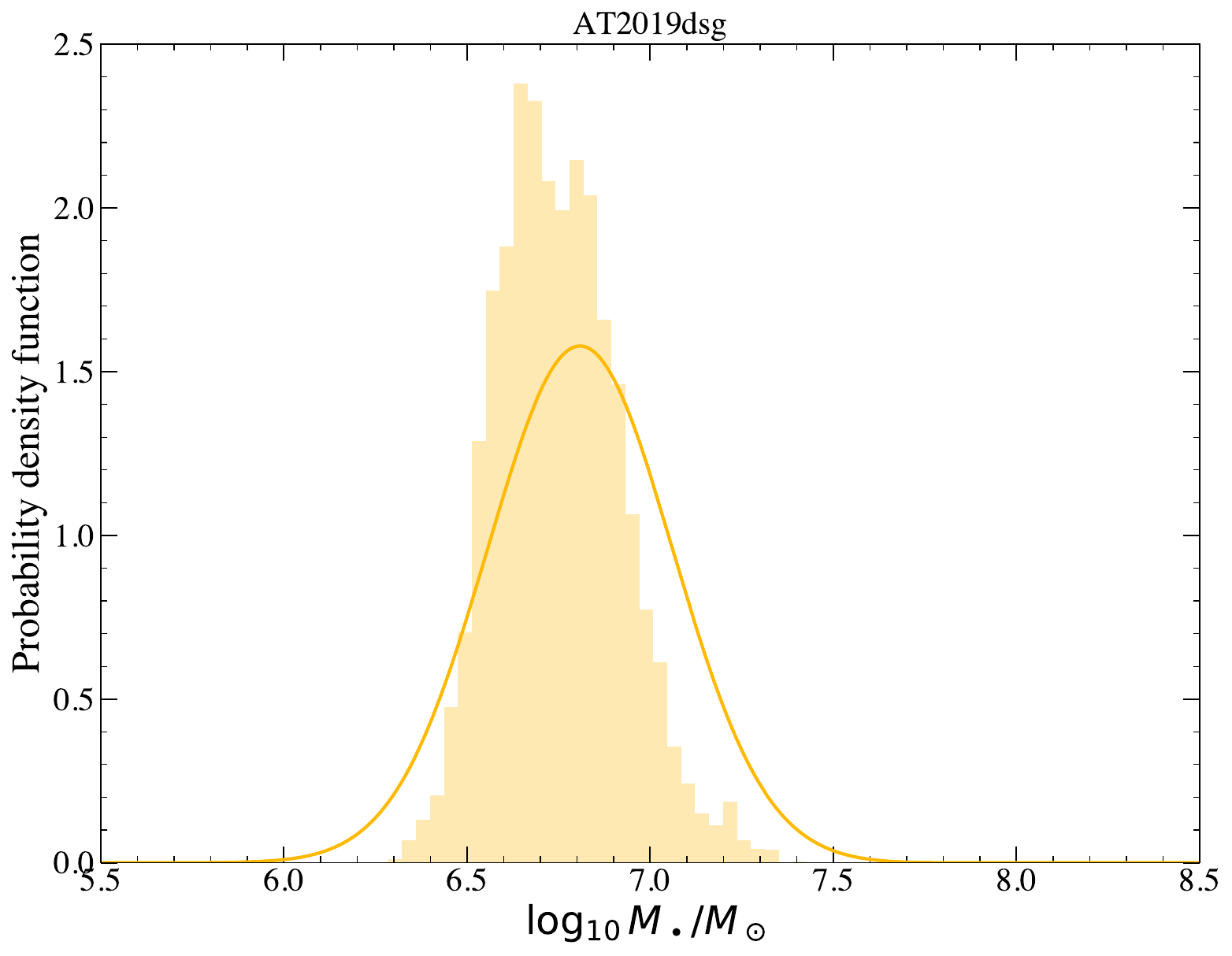}
    \includegraphics[width=0.35\linewidth]{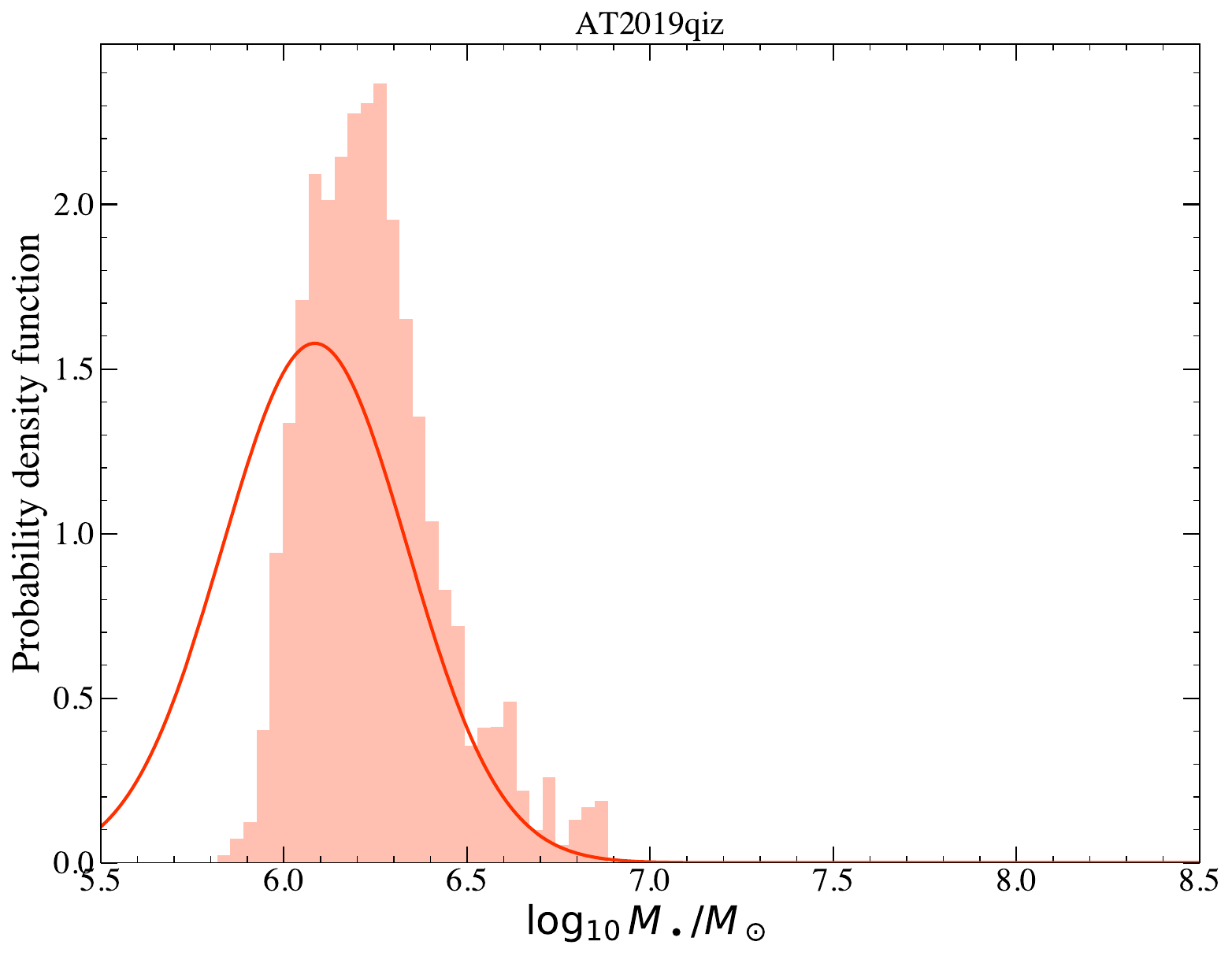}
    \includegraphics[width=0.35\linewidth]{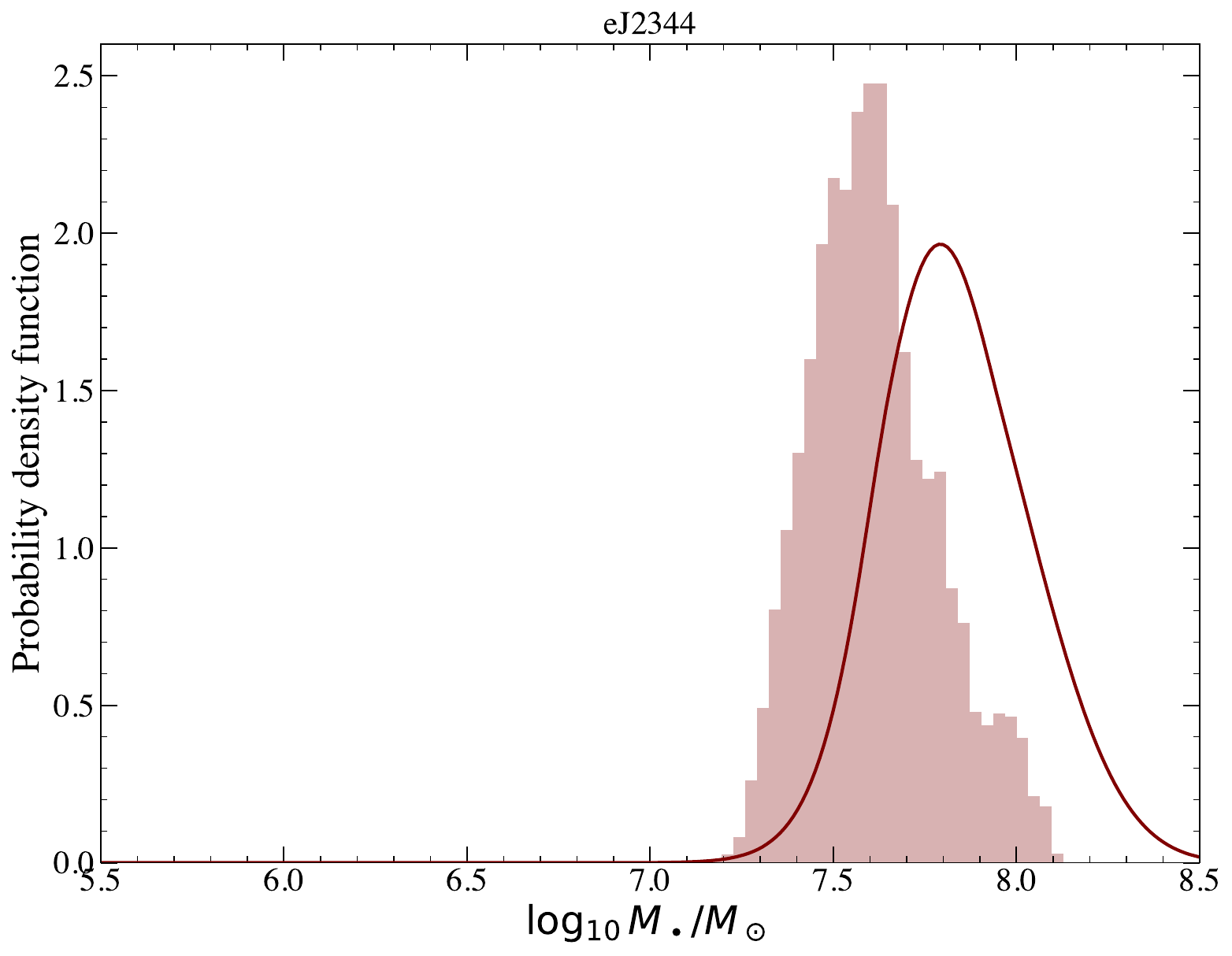}
    \caption{A comparison between the mass posteriors of sources fit with {\tt TDEFLARE} and the more physically motivated {\tt FitTeD} code. All sources have black hole mass values which are consistent at $1\sigma$. The {\tt TDEFLARE} code provides a rapid means to infer the black hole mass in a TDE, but lacks interpretability (and more detailed physical understanding) provided by codes which contain more physics.   }
    \label{fig:compare}
\end{figure*}

AT2022dbl is the ideal TDE to use as a test here, as it has been observed to undergo two flares \citep{Yao24dbl, Lin24}, and has a published $M_\bullet-\sigma$ mass \citep{Makrygianni25}. In Figure \ref{fig:22dbl} we show (in the top row) the two flares observed from AT2022dbl (data from {\tt manyTDE}). We split the two flares and fit them independently, with parameter posteriors shown in the middle row. All black hole mass posteriors are shown in the bottom panel, with inference from the first flare shown by dashed curves, and the second flare by solid curves. All are consistent at $1\sigma$. They are also consistent with the $M_\bullet-\sigma$ value. We conclude that {\tt TDEFLARE} also works for partial TDEs. AT2022dbl also highlights how little data is needed for mass inference using {\tt TDEFLARE} (i.e., the second flare), a question we shall return to shortly.

\subsection{{\tt TDEFLARE} versus {\tt FitTeD}}
The first two examples (AT2019dsg and eJ2344) showed that, once properties of the optical flare can be constrained from the data, the posterior distributions of the black hole mass are consistent with models which contain significantly more physics. 

In this section we highlight that this appears to be a generic property of these solutions. We show this by plotting the posterior of the {\tt TDEFLARE} model (solid curves) once all data is modeled and the Hills physics is taken into account,  compared to the posterior mass of a full {\tt FitTeD} model (which generally includes evolving X-ray data and a full model of the disk). 

In Figure \ref{fig:compare} we compare these two mass distributions for 8 TDEs for which a full {\tt FitTeD} analysis has been performed (a generally much more expensive computational process; these fits and a discussion are presented in Goodwin \& Mummery, submitted.). The mass distributions are all consistent at $1\sigma$. Of course, what one gains in modeling simplicity (and computational speed) one loses in interpretability, as {\tt FitTeD} also constrains the disk mass, the black hole spin, the viscous timescale, etc. Further, {\tt FitTeD} also allows (e.g.,) the accretion rate onto the black hole to be constrained at all times, of interest to various fundamental questions of disk theory. None of the above can be inferred from {\tt TDEFLARE}, but the mass inference is consistent between both approaches. 

\subsection{How much data is needed?}
Detecting a optical/UV plateau is observationally expensive, as it requires long baselines, stacking of data and in an ideal world UV observations. Constraining the peak spectral luminosity and radiated energy requires (in principle) two data points (although of course $N>2$ is preferable). It is worth asking by how much {\tt TDEFLARE} mass inference varies as a function of time since optical peak. 
\begin{figure*}
    \centering
    \includegraphics[width=0.35\linewidth]{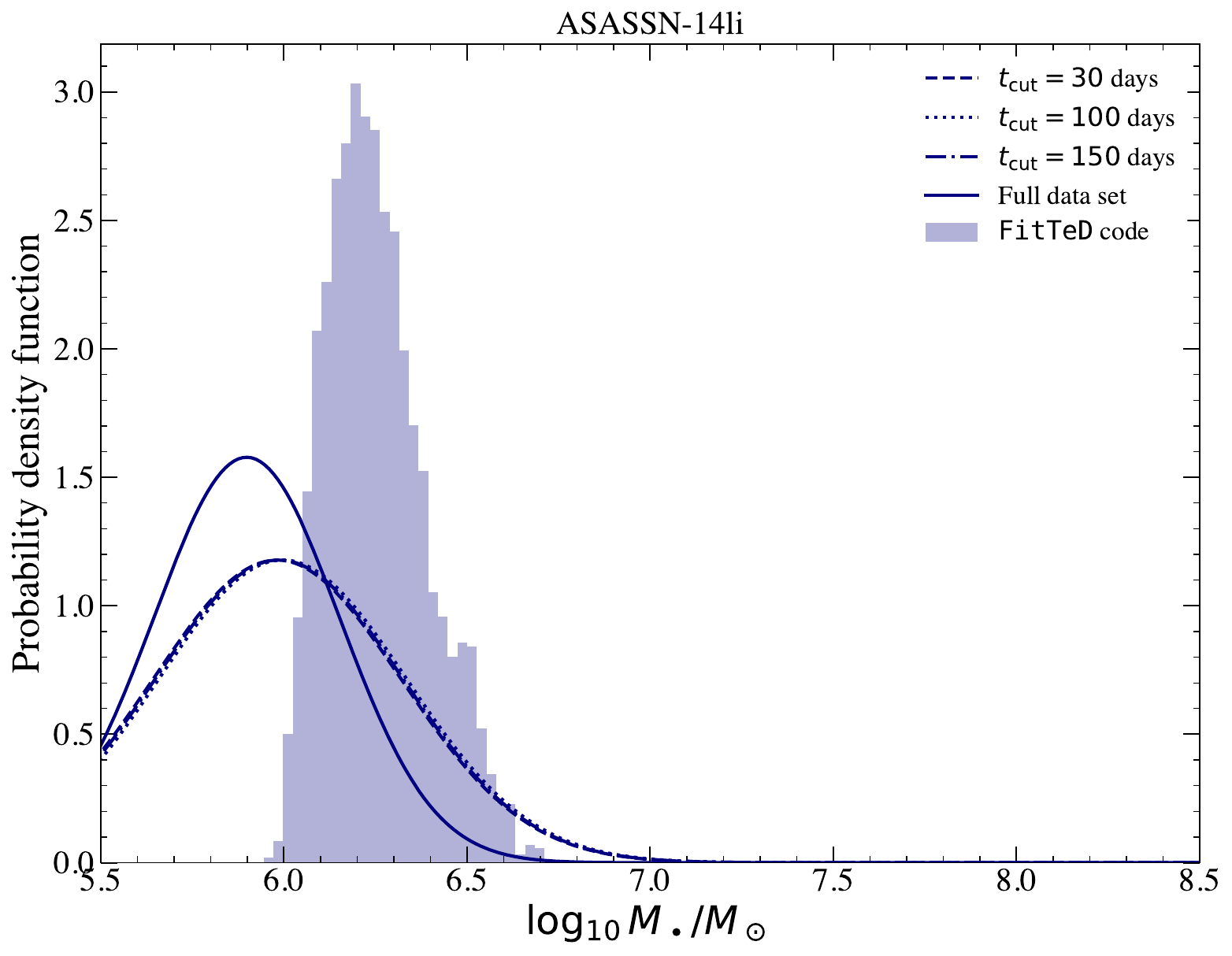}
    \includegraphics[width=0.35\linewidth]{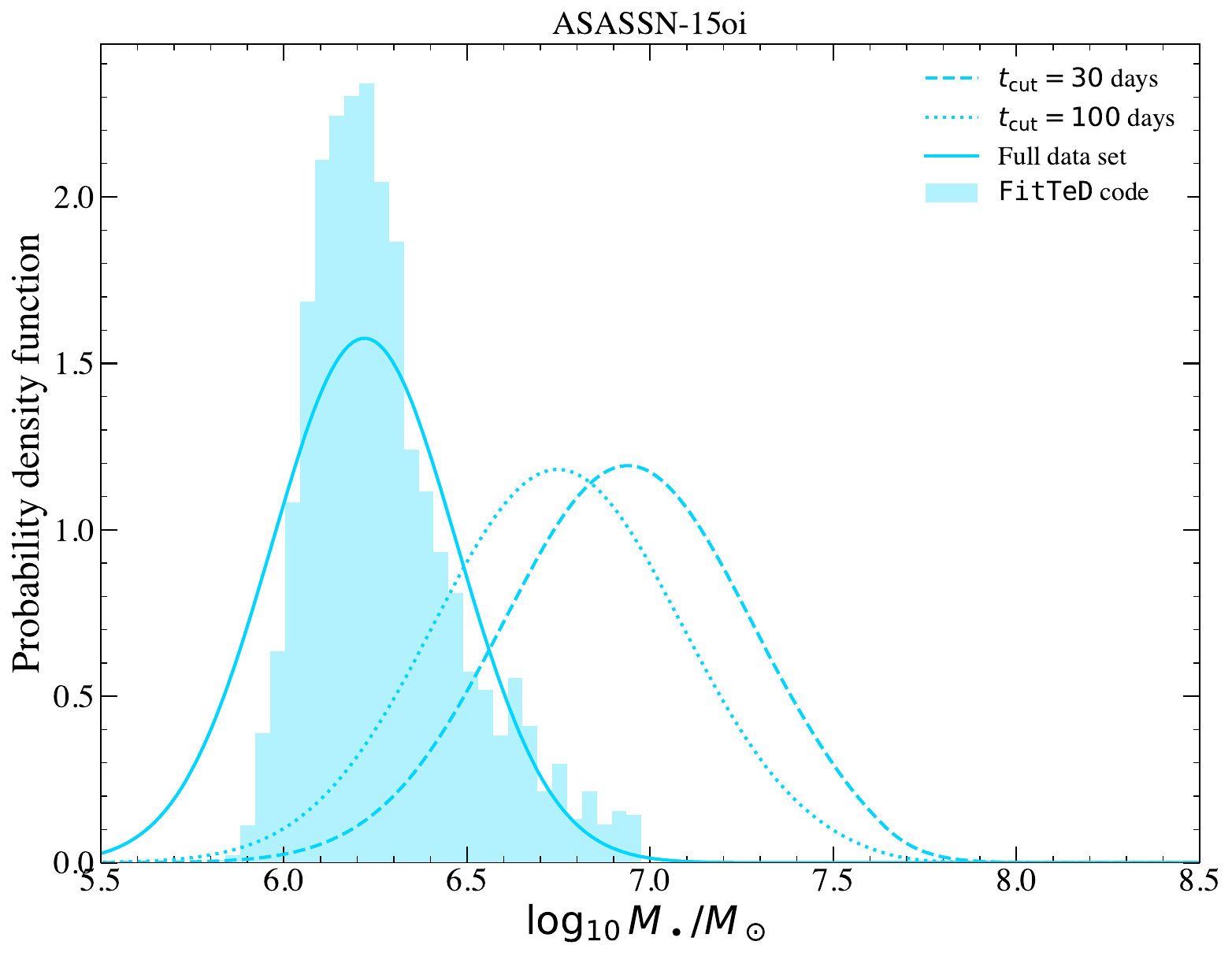}
    \includegraphics[width=0.35\linewidth]{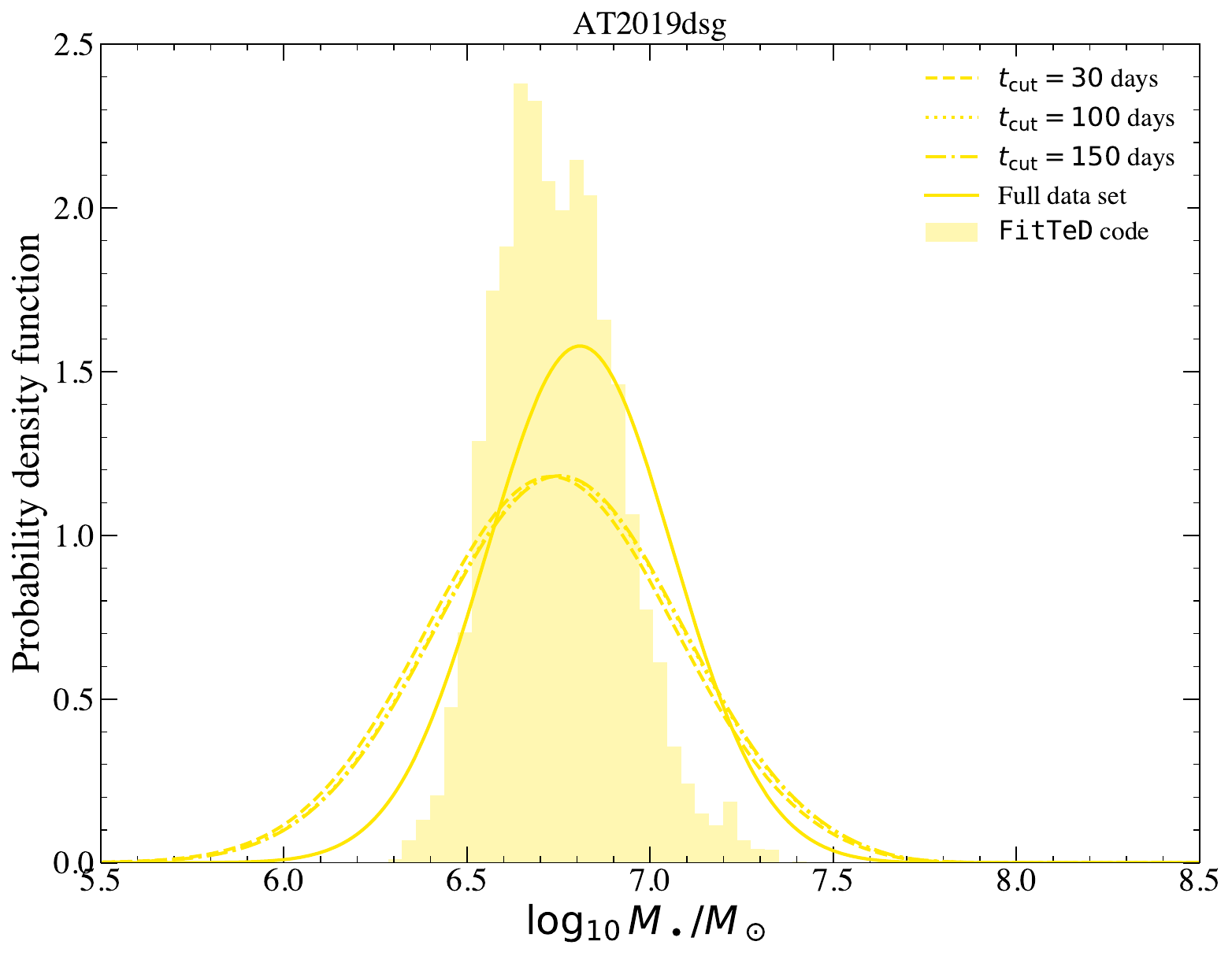}
    \includegraphics[width=0.35\linewidth]{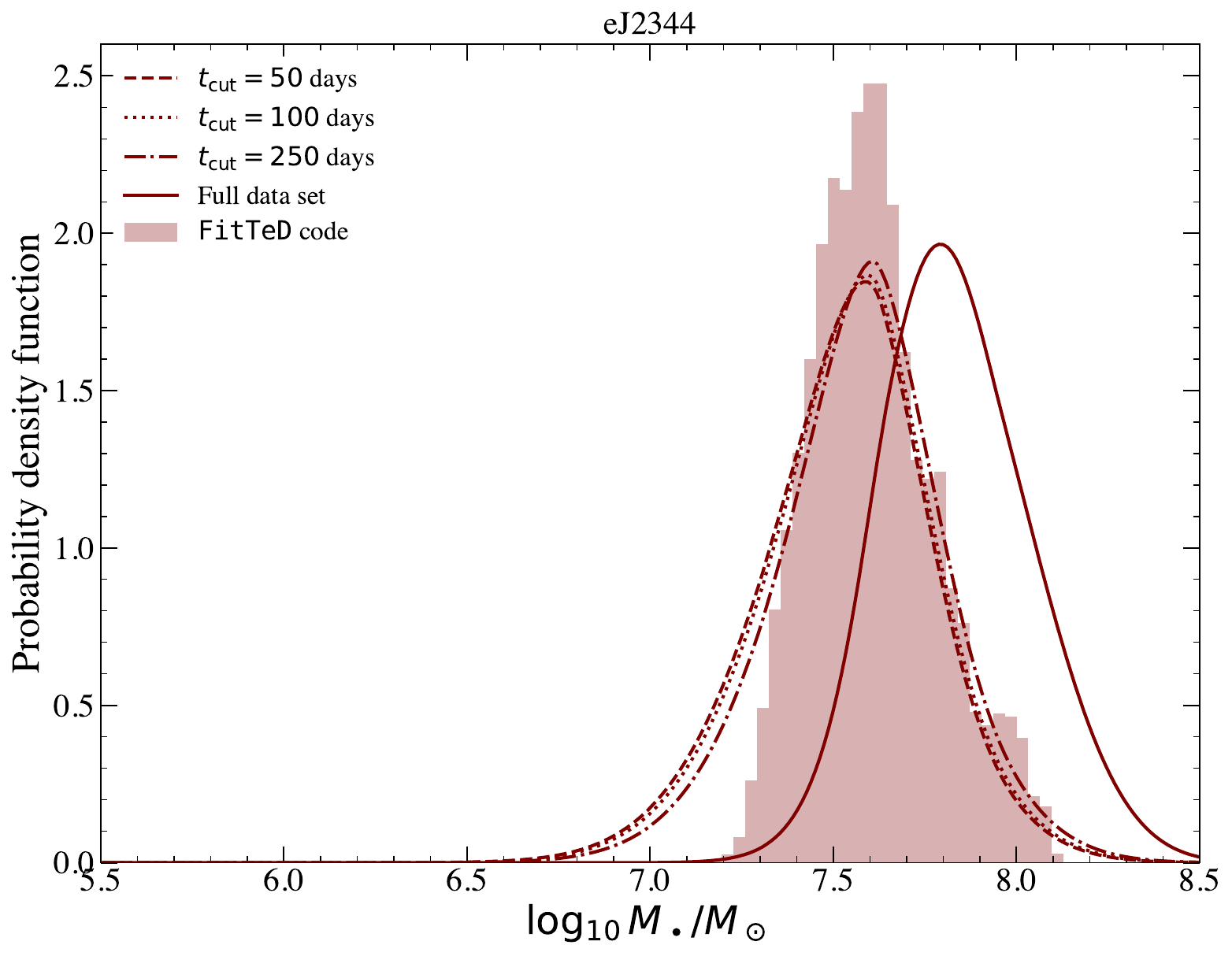}
    \caption{The impact of collecting more data post peak on black hole mass inference with {\tt TDEFLARE}. In this plot we compute the mass posteriors as a function of time since peak (defined by deleting all data with $t-t_{\rm peak}>t_{\rm cut}$). Waiting for the plateau generally shrinks the uncertainty on the mass, but causes minimal absolute shifts in the inferred masses. Exceptions do exist, including ASASSN-15oi (see text for more discussion). Waiting for the plateau of course allows more physically motivated models to be fit to the data (e.g., {\tt FitTeD} and {\tt kerrSED}), but rapid -- and reasonably accurate -- black hole estimation is possible with {\tt TDEFLARE} from relatively limited optical/UV datasets. }
    \label{fig:cut}
\end{figure*}

In Figure \ref{fig:cut} we show the black hole mass posteriors as a function of the time at which we stop including data (i.e., we delete all data for times $t-t_{\rm peak}>t_{\rm cut}$) for four TDEs from different mass scales. We  compute the black hole mass posteriors from the posteriors on $L_{\rm pk}$ and $E_g$ (as it is obviously not possible to constrain the plateau luminosity from this early time data). Generically, only a handful of data points are required to determine a constraint on the TDEs black hole mass from early time data. As a rule, this posterior does not vary much as more data is added, until the plateau becomes robustly detectable at which point the uncertainty in the mass drops (as more information is available), and there is often a shift in the posterior median. 

We include ASASSN-15oi in Figure \ref{fig:cut} as a note of caution here however, as this is one of a small number of sources which show strongly changing posteriors even before the plateau is detected. The reason for this is that ASASSN-15oi was fit with only UV data at early times, and so the values of the peak luminosity and radiated energy {\it in the $g$-band} are poorly constrained (as they are more sensitive to the temperature, a nuisance parameter). As LSST will always record $g$-band data for future TDEs, we do not expect this to be much of a problem going forward, but it is essential that this possible effect is taken into account. 

\section{Discussion}\label{discussion}
The previous section demonstrated the use of {\tt TDEFLARE} on the single source level. More interesting will be the population-level inference of black hole masses that can be performed with the large sample of TDEs discovered by LSST. In this section we highlight some of the physics that can be done already, and discuss some biases present in the answers to interesting astrophysical questions. 

\subsection{Galactic scaling relationships}
Do TDEs follow standard galactic scaling relationships extrapolated down to lower masses? While initial attempts to answer this question with the early-time optical flare from TDEs failed to find a statistically significant correlation between galaxy properties and TDE-model black hole masses \citep{Ramsden22, Hammerstein23, Guolo25c}, it is now known that models which use late-time optical/UV data \citep{Mummery_et_al_2024, MummeryVV25, Ramsden25} or joint optical/UV-X-ray data \citep{Guolo25c} recover these correlations at high ($>5\sigma$) significance. 

Showing that {\tt TDEFLARE} also recovers known galactic scaling relationships at high significance is therefore not new, but it does allow the uncertainty on each black hole mass to be reduced (from that inferred from the plateau method), and the sample size to expanded (compared to that using full SED/light curve fitting). 

\begin{figure*}
    \centering
    \includegraphics[width=0.48\linewidth]{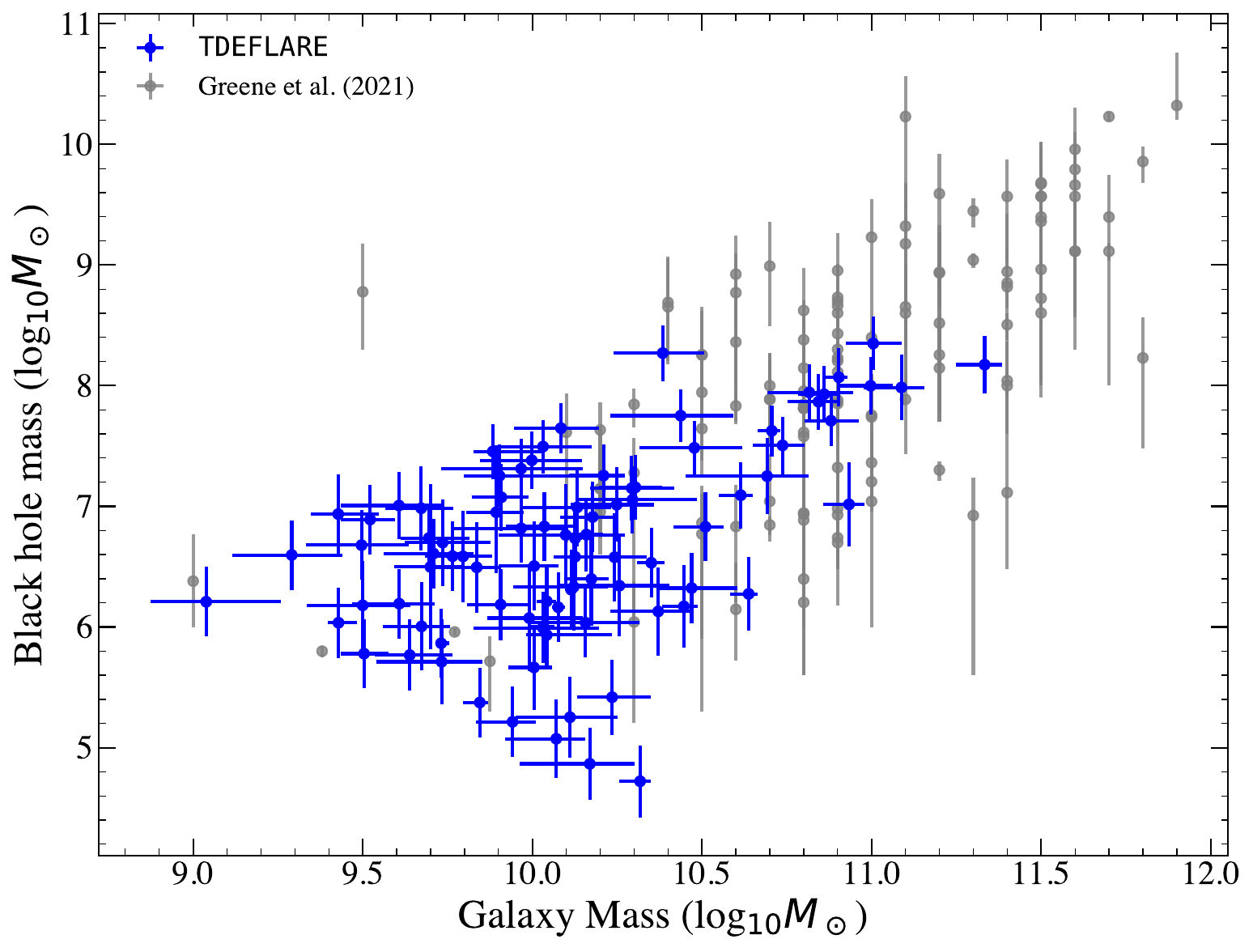}
    \includegraphics[width=0.48\linewidth]{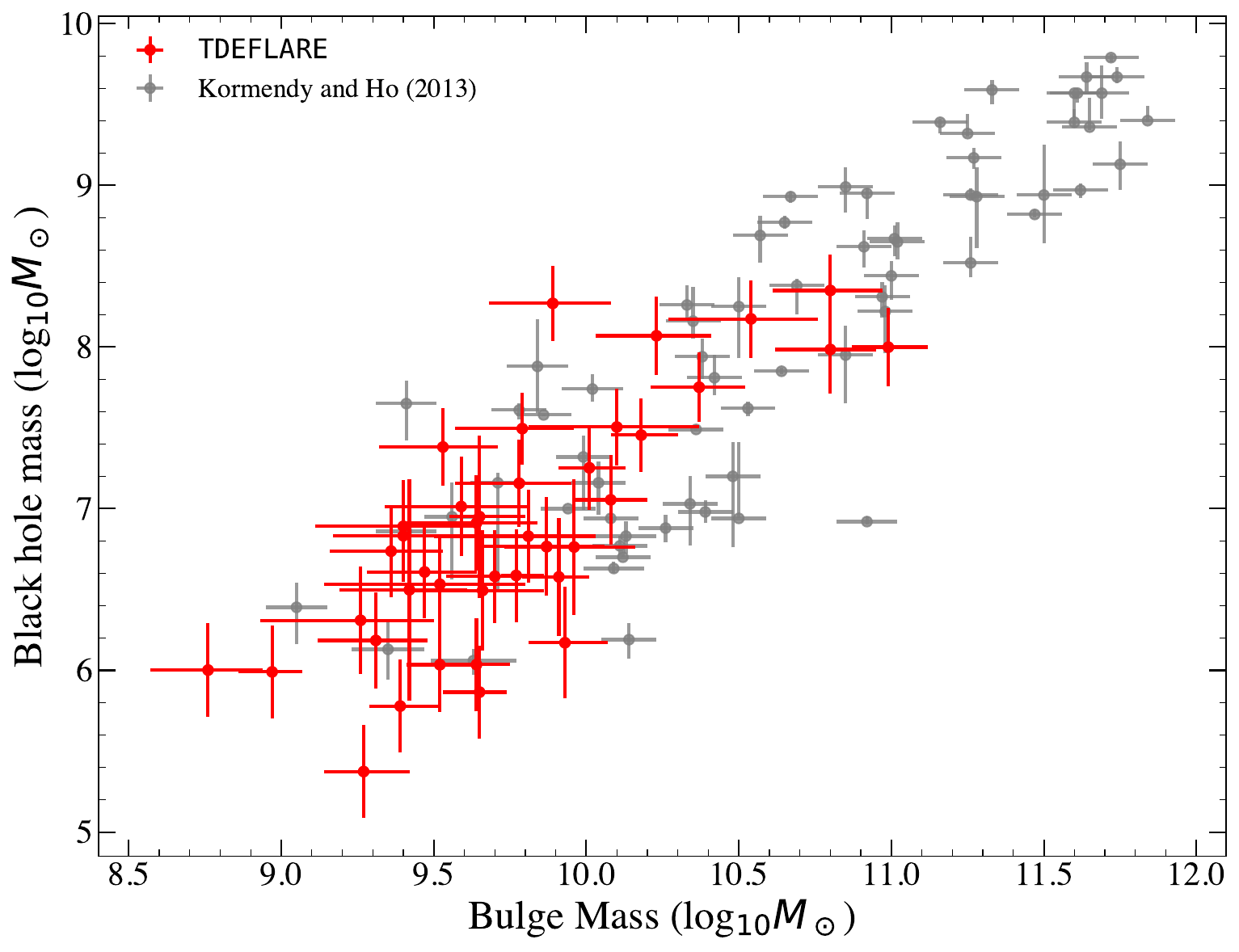}
    \includegraphics[width=0.55\linewidth]{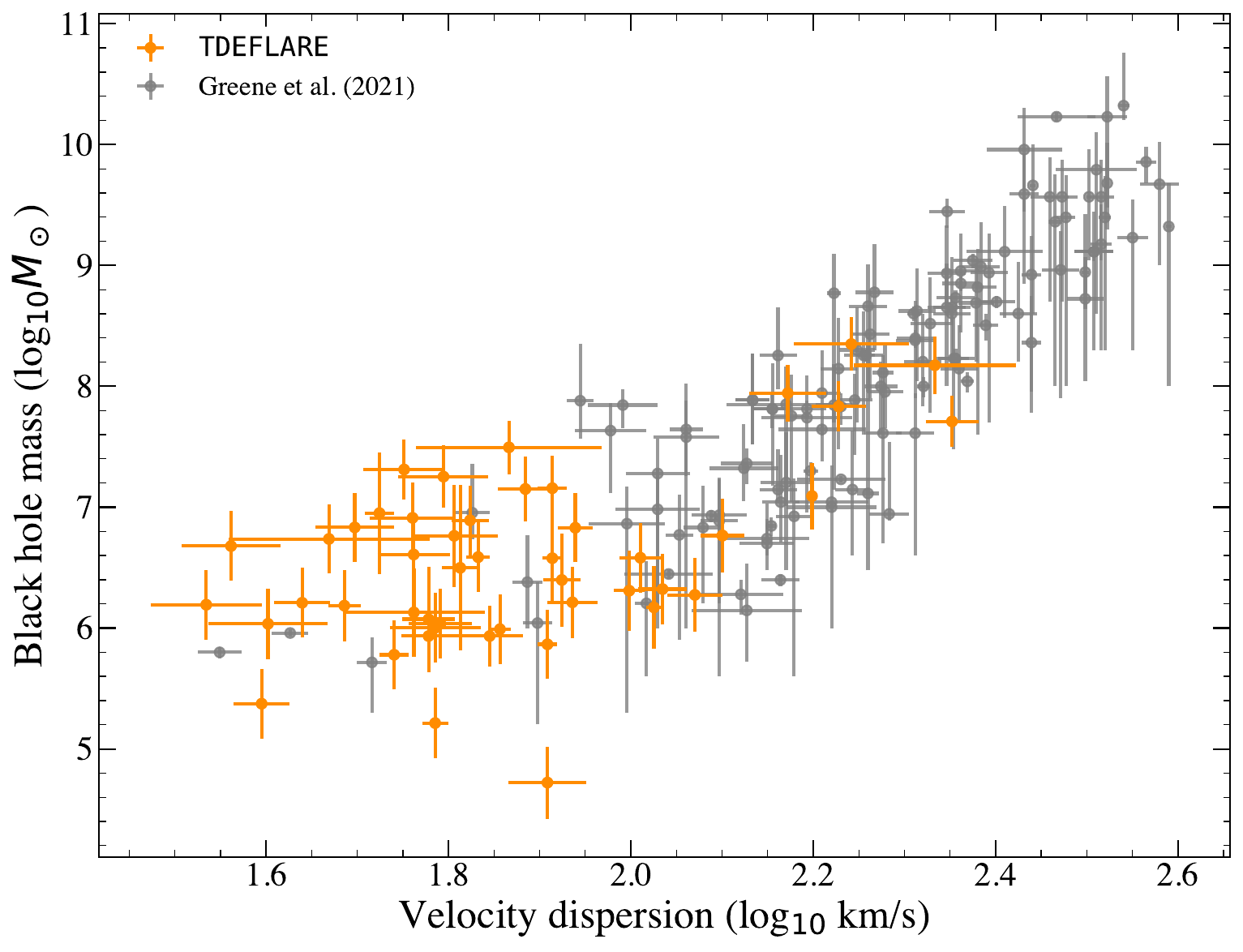}
    \caption{Fitting the $N=89$ TDEs from the {\tt manyTDE} data base with {\tt TDEFLARE} allows $88$ TDEs to be added to the black hole -- galaxy mass correlation (upper left), $40$ sources to be added to the black hole mass -- bulge mass correlation (upper right), and $47$ sources to be added to the black hole mass -- velocity dispersion correlation (lower). The TDEs added to these scaling relationships by {\tt TDEFLARE} have masses which are consistent with, but are at systematically lower masses than, those black holes added by other techniques.    }
    \label{fig:galaxies}
\end{figure*}

In Figure \ref{fig:galaxies} we show the results of fitting {\tt TDEFLARE} to the full {\tt manyTDE} sample (and eJ2344). For the following TDEs: AT2019mha, AT2019lwu, AT2019teq, AT2020pj, AT2019vcb, AT2020ddv, AT2020ocn, AT2020mbq, AT2021jjm, AT2021uvz, AT2022exr, AT2023rvb, AT2023mfm and AT2021utq the plateau luminosity $L_P$ was poorly constrained, so mass inference only used $L_{\rm pk}$ and $E_g$. For AT2022dbl we used the mass implied by fitting the first flare (which makes no difference). All galaxy properties are taken from {\tt manyTDE}, except the bulge masses which were taken from \cite{Ramsden25}. 

Fitting the $N=89$ TDEs from the {\tt manyTDE} data base (and eJ2344) with {\tt TDEFLARE} allows $88$ TDEs to be added to the black hole -- galaxy mass correlation (upper left), $40$ sources to be added to the black hole mass -- bulge mass correlation (upper right), and $47$ sources to be added to the black hole mass -- velocity dispersion correlation (lower). The TDEs added to these scaling relationships by {\tt TDEFLARE} have masses which are consistent with, but are at systematically lower masses than, those black holes added by other techniques.  Each of the three scaling relationships found from {\tt TDEFLARE} has extremely strong statistical significance ($>5\sigma$). 

\subsection{The TDE black hole mass function}
Individual black hole masses, and their relation with galaxy properties, are of course of interest to TDE science. However, perhaps the most interesting question which can be probed by observations of TDEs is what is the intrinsic mass distribution of low-mass black holes? 

\begin{figure*}
    \centering
    \includegraphics[width=0.45\linewidth]{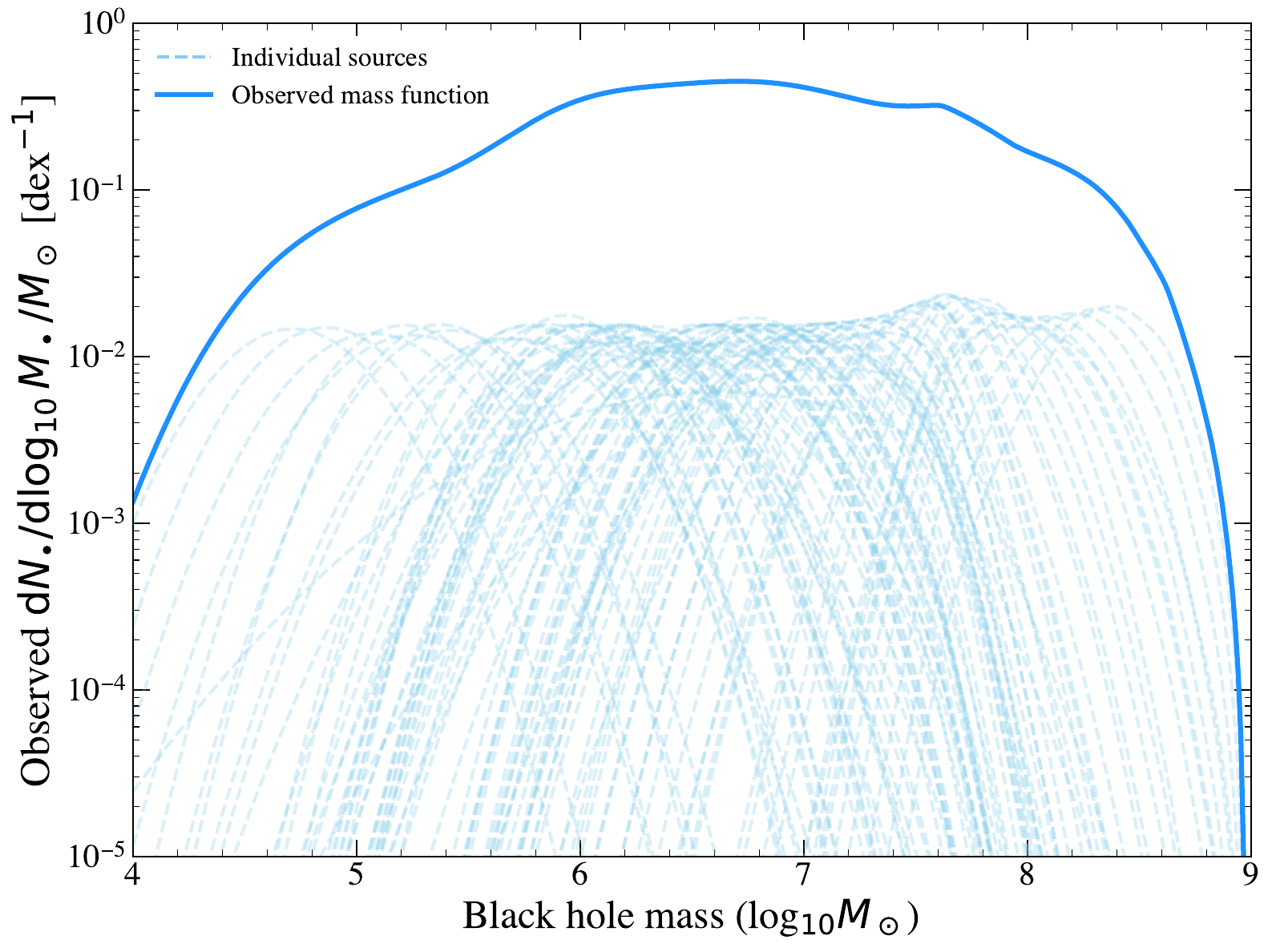}
    \includegraphics[width=0.45\linewidth]{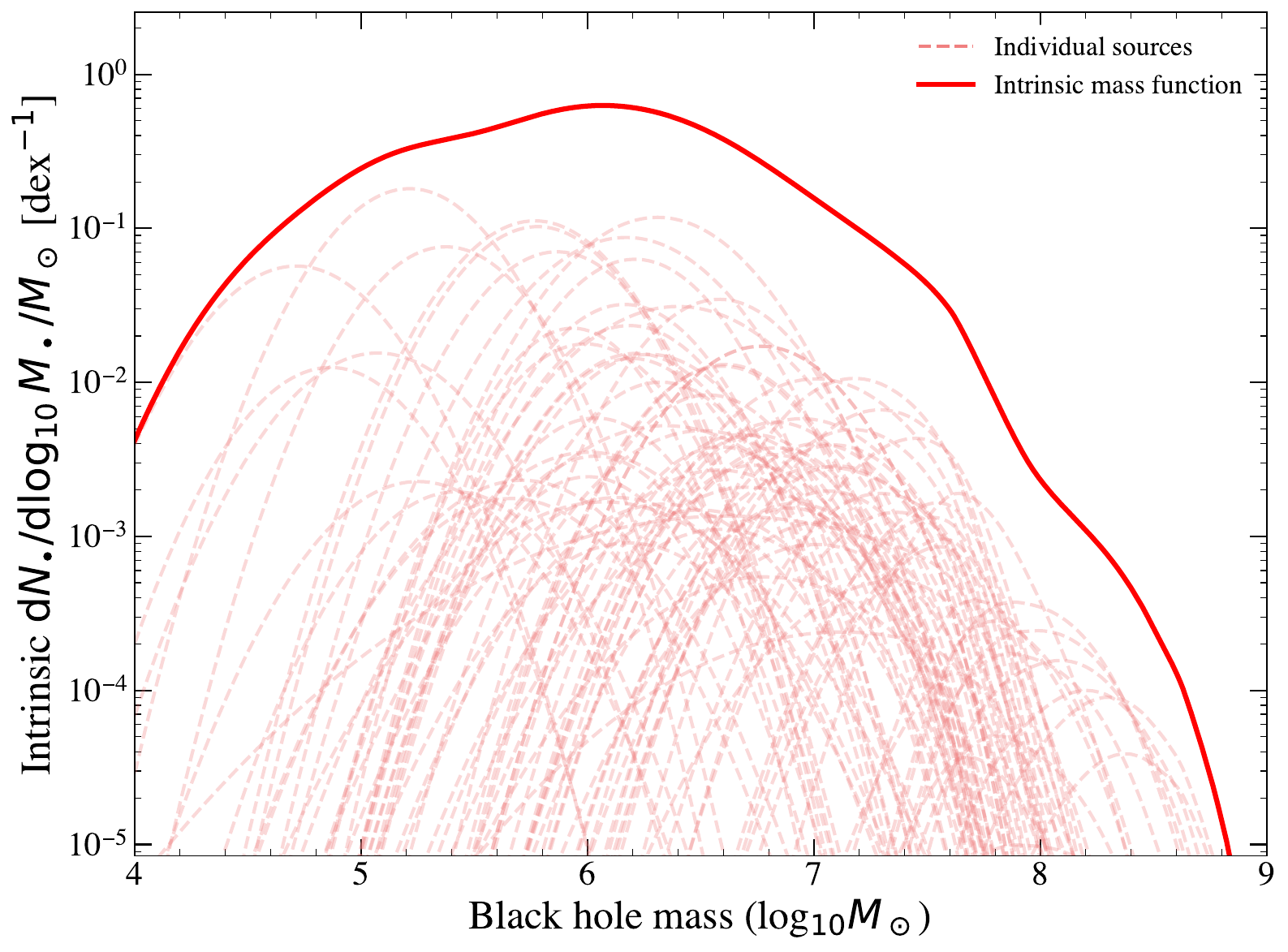}
    \caption{The black hole mass function implied by the 89 TDEs modeled in this work. On the left we display the {\it observed} black hole mass function (normalised to have an integral of 1) as a function of black hole mass (solid curve), with individual sources displayed by dashed curves. The observed black hole mass function is relatively flat, and extends up to high  masses before dropping off super-exponentially, implying that the observed TDE population includes a sub-population of rapidly rotating high-mass black holes.  The right panel shows the (more interesting) {\it intrinsic} black hole mass function (normalised to have an integral of 1) in our TDE population (i.e., the black hole mass function once the differing observing volumes of different sources are corrected for). This is peaked at around $M_\bullet \sim 10^6 M_\odot$, decays roughly as a power law for $M_\bullet \sim 10^6-10^8M_\odot$, before rapidly dropping to zero for $M_\bullet\geq 10^8M_\odot$. Both distributions are consistent with the  calculations of \citealt{MummeryVV25}. Note that the two distributions both drop to zero relatively rapidly below $M_\bullet\lesssim 10^5M_\odot$ -- this could be entirely a Malmquist bias and should not be interpreted as the non-existence of IMBHs.   }
    \label{fig:mass_function}
\end{figure*}

Using {\tt TDEFLARE} this intrinsic mass function is simple to compute. Most simple to compute is the {\it observed} black hole mass distribution, which is simply the sum of all of the individual black hole mass posteriors of our sample ($N=89$). This is shown in the left hand panel of Figure \ref{fig:mass_function}. Each individual posterior is shown by a dashed curve (note the heights of each posterior are comparable because the uncertainty on each mass are broadly the same across the population), while the sum is shown by the solid curve. The absolute normalisation of this distribution is not constrained by {\tt TDEFLARE} (and can be related to the absolute rate of TDEs), so we normalize the curve to have an integral of 1. The observed black hole mass function is relatively flat, with a peak at $\log_{10}M_\bullet/M_\odot \approx 6.5$ (the most common  observed TDE black hole mass), before extending up to high  masses and thereafter dropping off super-exponentially. The extension above $10^8M_\odot$ implies that the observed TDE population includes a sub-population of rapidly rotating high-mass black holes.  

To go from an observed mass function to an intrinsic mass function, one must take into account the different observing volume available to each TDE. Put simply, intrinsically more luminous TDEs will be over represented in our (flux limited) sample, and if they systematically have certain masses (they do -- higher masses) they will cause the observed mass function to deviate from the intrinsic mass function. To do this in a truly careful manner one should take into account (i) the different selection functions of different observational surveys (something especially relevant for combined samples like that used here), (ii) the spectroscopic completeness of surveys as a function of magnitude, and (iii) the spectral shape of TDEs and the limiting fluxes of surveys {\it as a function of wavelength}. This final point is important for objects discovered at cosmological distances, as one effectively samples different rest frame frequencies for sources at different red shifts $z$. 

For this first analysis we shall neglect these complications (see \citealt{Yao23} for a detailed discussion with a smaller sample size). The vast majority of our sample are ZTF TDEs discovered at low redshift, and so this is unlikely to be a significant simplification. With {\tt TDEFLARE}, it is relatively easy to go from an observed to intrinsic black hole mass function, as we are also modeling the luminosity with which our survey finds TDEs (the $g$-band luminosity), and so the {\it relative} observable volume of different TDEs can be easily estimated. The absolute volume ${\cal V}$ for which a TDE could be detected by an  optical survey with a  flux detection limit, $F_{\rm lim}$, can be simply estimated from the maximum distance out to which we could have found that TDE (i.e., the distance at which $F_{\rm obs}=F_{\rm lim}$). In a Newtonian cosmology (which is a reasonable approximation for nearly all tidal disruption events which are found at $z < 0.1$), the maximum distance out to which a source with given peak luminosity  $L_{{\rm pk}}$ can be observed is 
\begin{equation}
    D_{\rm max} \propto \sqrt{L_{{\rm pk}}} %\sqrt{L_{{\rm peak}} \over 4 \pi F_{\rm lim}}, 
\end{equation}
and therefore an estimate of the volume out to which each tidal disruption event can be detected is 
\begin{equation}
    {\cal V} \propto D^3_{\rm max} \propto \left(L_{{\rm pk}}\right)^{3/2} . 
\end{equation}
Therefore our intrinsic TDE mass function is simply the sum of the individual mass posteriors ($p(M_{\bullet, i})$), with each TDE reweighted by a relative volume
\begin{align}
    p_{\rm intrinsic}(M_{\bullet, i}) &= p(M_{\bullet, i}) ({\cal V}_0/{\cal V}_i) \nonumber \\ &= p(M_{\bullet, i}) (L_0/{L}_{{\rm pk},i})^{3/2},
\end{align}
where the reference volume/luminosity ${\cal V}_0/L_0$ are arbitrary (as we will renormalize the final distribution anyway). As every TDE has a measured value of $L_{{\rm pk}, i}$, this is a trivial calculation to perform. 

The right panel shows this (more interesting) {\it intrinsic} black hole mass function (again normalised to have an integral of 1) in our TDE population (i.e., the black hole mass function once the differing observing volumes of different sources are corrected for). This is peaked at around $M_\bullet \sim 10^6 M_\odot$, decays roughly as a power law for $M_\bullet \sim 10^6-10^8M_\odot$, before rapidly dropping to zero for $M_\bullet\geq 10^8M_\odot$.

We note that both of the mass distributions inferred in this work (the observed and intrinsic) are consistent with the first principle calculations of \cite{MummeryVV25}, see their figure 11.  

\subsection{Malmquist-Hills bias}
While the {\tt TDEFLARE} model recovers known galactic scaling relationships at extremely high statistical significance ($>5\sigma$), a result in common with other accretion-based models \citep{Mummery_et_al_2024, Guolo25c}, it is clear (even visually) that the $M_\bullet-\sigma$ relationship followed by TDEs is flatter than that which would be inferred from extrapolating the high-mass sample down to lower masses (Figure \ref{fig:galaxies}). (A similar result was found in \cite{Ramsden25} for the bulge mass relationship, where the following bias was also discussed.) 

\begin{figure*}
    \centering
    \includegraphics[width=0.55\linewidth]{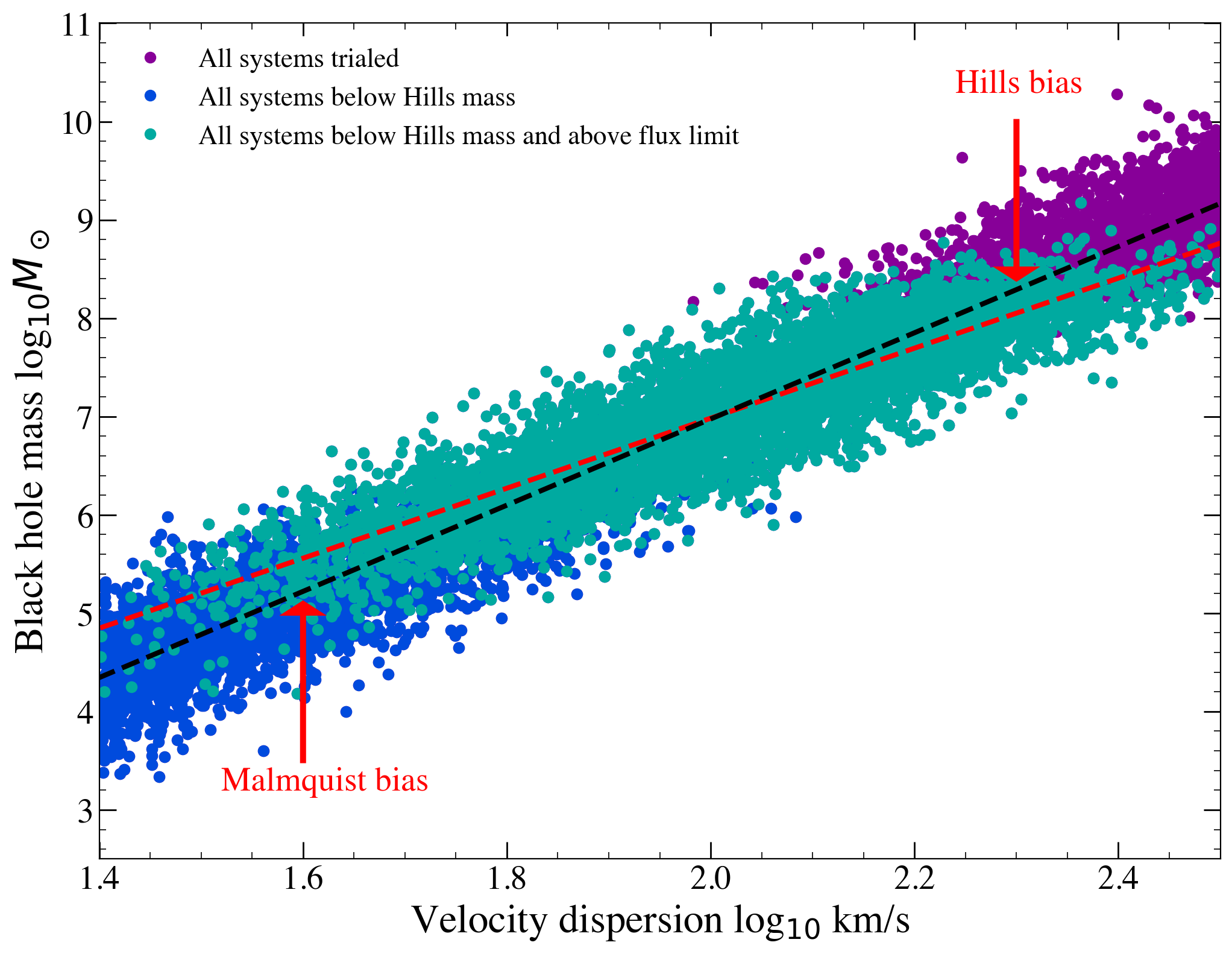}
    \includegraphics[width=0.43\linewidth]{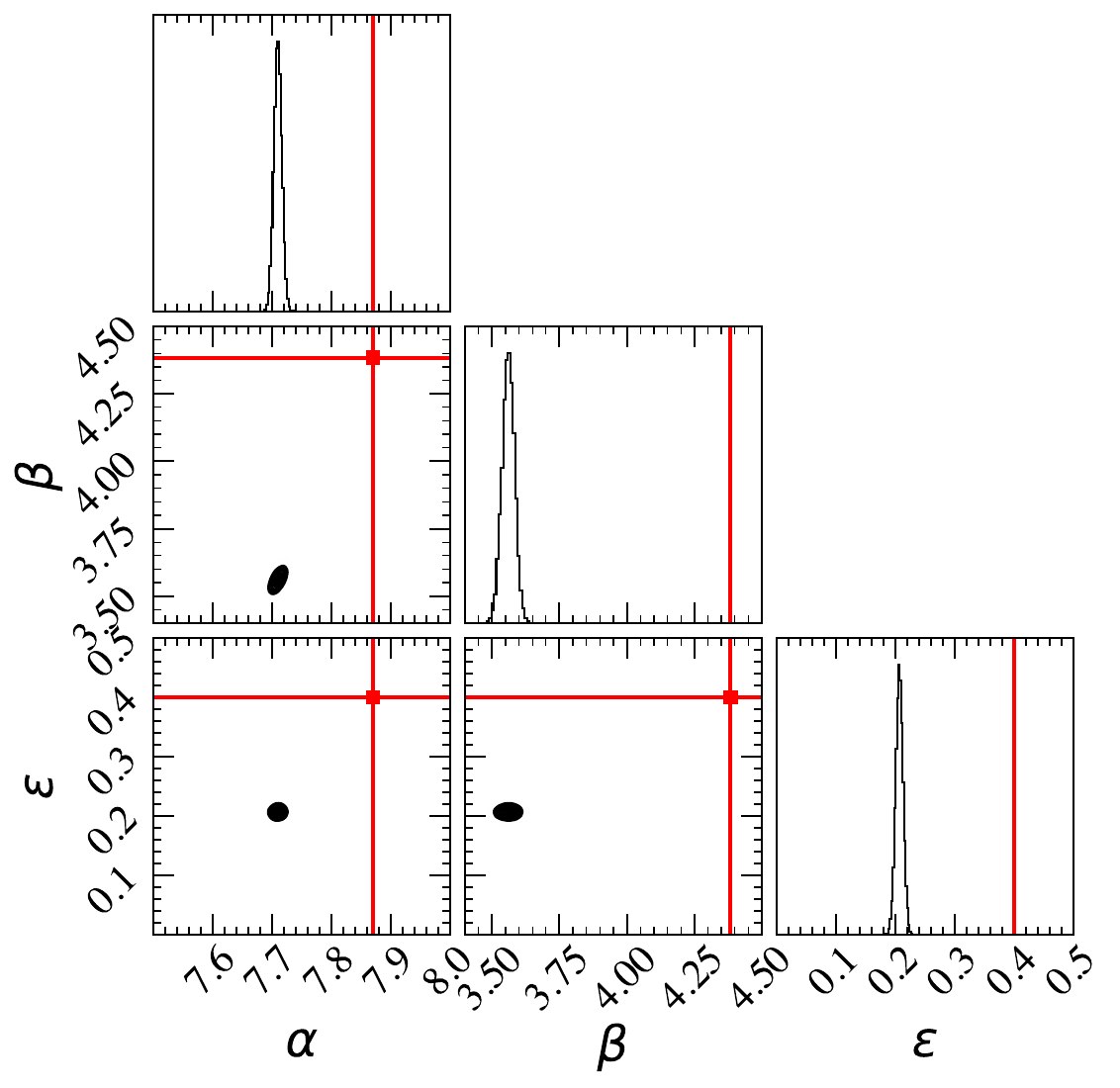}
    \caption{Population-synthesis of optical TDEs, highlighting the Malmquist-Hills bias which acts to flatten TDE-only scaling relationships from their true values. In the left hand panel we show a sample of $10,000$ TDEs, with velocity dispersions uniformly distributed (on a log scale) between 1.4 and 2.5, for which a black hole mass can be assigned from a ``true'' scaling relationship (black dashed curve with 0.4 dex of Gaussian scatter). At the high $\sigma$ end many systems lie above the Hills mass, suppressing the $M_\bullet-\sigma$ relationship (``Hills bias''). At the low mass end only over-massive (for their $\sigma$) systems lie above the ZTF flux limit (``Mlamquist bias''). These two effects combine to flatten the relationship (red dashed curve). The posterior distribution of a power-law fit to the TDEs is shown in the right panel (black posteriors), with the ``truth'' (injected values) shown by red points/lines. The power-law index is reduced from $\beta_{\rm true} = 4.4$ to $\beta_{\rm measured}\approx 3.5$.  }
    \label{fig:malm-hills}
\end{figure*}

There are dynamical reasons to suspect that the $M_\bullet-\sigma$ relationship might flatten out at low masses \citep[see e.g.,][for a review]{Greene20}, but before one may claim a detection of such a feature one must be careful to take account of two natural biases in any TDE relationship, which we shall show here will always flatten an inferred galaxy scaling relationship. We discuss this dual bias with a particular focus on the $M_\bullet-\sigma$ relationship. 

These two biases are the following: at high black hole mass scales (and fixed galaxy property), a TDE is much more likely to occur for a black hole which has a mass which is {\it lower} than would be expected for its galactic scaling relationship \citep[a result also discussed in][]{Ramsden22}. The reason for this is that these lower mass black holes have, for a given star, a tidal radius further from the event horizon and are therefore more likely to produce observable emission from a TDE, rather than swallow a star whole. We dub this the ``Hills bias''.  To be explicit, a TDE observed in a galaxy with velocity dispersion $\sigma = 250$ km/s is much more likely to have mass $M_\bullet=10^8M_\odot$ than $M_\bullet=10^9M_\odot$. This means, that at the high mass end of a given galactic scaling relationship, TDEs will be systematically under-massive. 

At low masses, however, TDEs will be systematically over-massive for a given galaxy property. This is simply a result of the luminosity of TDEs increasing with black hole mass (Figure \ref{fig:basis}). This is just a classic Malmquist bias, in a flux limited survey we always find more over-luminous (i.e., over-massive in the case of TDEs) sources at a fixed value of any other property. This pushes up black hole masses at the low galaxy property end. 

It is simple to see that by pushing masses down at the high end, and masses up at the low end, this combined ``Malmquist-Hills'' bias will flatten a given scaling relationship. Caution must therefore be taken before interpreting a given observed flattening of a scaling relationship with TDE data. 

To make this point quantitative, we perform a simple population-synthesis calculation (Figure \ref{fig:malm-hills}). We sample velocity dispersion measurements according to $\log_{10} \sigma \sim {\cal U}(1.4, 2.5)$ where ${\cal U}$ is the uniform distribution. We then assign to each system a value the black hole mass from this sampled $\sigma$ using the \cite{Greene20} scaling relationship, namely 
\begin{equation}
    \log_{10} M_\bullet/M_\odot = 7.87 + 4.39 \log_{10} \sigma/160\, {\rm km/s}, 
\end{equation}
with a (normally distributed) intrinsic scatter of $\epsilon = 0.4$ dex. Every trialed TDE system is plotted as purple points in Figure \ref{fig:malm-hills}. For each TDE we then compute a Hills mass by sampling a random star, a random incoming stellar orbital inclination, and a random black hole spin (using the default {\tt tidalspin} distributions, \citealt{Mummery24}), and keep only those systems with $M_{\bullet, i} < M_{\rm Hills}$. These systems are shown by dark blue points in Figure \ref{fig:malm-hills}, where we see the systematic suppression of masses at the high $\sigma$ end. 

We then use the empirical $g$-band peak luminosity scaling relationship (Figure \ref{fig:basis}) to assign each of these TDEs a peak $g$-band luminosity (with intrinsic scatter of $\epsilon = 0.5$ dex). We assume that TDE host galaxies are uniformly distributed in a spherical volume around the observer, out to a maximum distance $D_{\rm max} = 500$ Mpc (i.e., each TDE is at a distance $D_i \sim D_{\rm max} \times ({\cal U}(0, 1))^{1/3}$), allowing us to compute a peak $g$-band flux for each TDE. 

We then keep only those TDEs with a peak $g$-band flux higher than the ZTF limiting magnitude of $m_{g} = 20.8$ (or a spectral flux limit of $F_{\rm lim} =  {1.74\times 10}^{{-28}}\text{\ erg\ s}^{{-1}}\text{\ cm}^{{-2}}\text{\ Hz}^{{-1}}$ at $\nu = 6\times 10^{14}$ Hz). This is obviously optimistic, as one would in reality need a flux significantly above the limit to confidently detect a TDE, but this process is only intended to highlight the broad statistical point. We also then optimistically assume that for every TDE in this sample with $F_i > F_{\rm lim}$ we can perfectly infer the mass. 

The systems that pass both the Hills mass and flux-limit constraints are shown by green points in Figure \ref{fig:malm-hills}, were we see that the flux limit induces a strong Malmquist bias at masses $\lesssim 10^5M_\odot$ (which explains the turnover in the TDE mass function at the low mass end, Figure \ref{fig:mass_function}). 

This causes a flattening of the inferred $M_\bullet-\sigma$ relationship, as can be seen by power-law fits to the TDE population (red dashed line shows the posterior median of the {\tt emcee} fit \citealt{EMCEE}), compared to the injected relationship (black dashed curve).  
In the right hand panel of Figure \ref{fig:malm-hills} we show the posterior distributions of a fit of the form 
\begin{equation}
    \log_{10} M_\bullet/M_\odot = \alpha + \beta \log_{10} \sigma/160\, {\rm km/s}, 
\end{equation}
to the TDE-only data (we include an intrinsic scatter of $\epsilon$ in the fit, which is to be compared to the value of 0.4 dex assumed in constructing the population). The injected value are shown by red points/lines in Figure \ref{fig:malm-hills}. The fitted relationship is significantly flatter $\beta \approx 3.5$, rather than $\approx4.4$ for the injected relationship. Significantly more data than might be naively expected will be required to robustly detect a flattening in $M_\bullet-\sigma$ with TDEs. The exact same caveats apply to $M_\bullet-M_{\rm bulge}$ and $M_\bullet-M_{\rm gal}$. 

Finally, we note a non-trivial quirk of these two biases -- the inferred scatter in the $M_\bullet-\sigma$ relationship of TDEs is $\sim $ half that of the injected relationship, a result of the dynamic range of the black hole masses being reduced. This of course assumed we could measure the black hole masses perfectly, and so will not trivially carry over to a more realistic population, but is worth keeping in mind in future work. 

We wish to stress one important, and as far as we can tell so far unappreciated, point here. What we have demonstrated here \citep[as has been discussed in previous works][]{Ramsden22, Ramsden25} is that TDEs are {\it biased} tracers of any given underlying galactic scaling relationship. What this means in practice is that if one believes a given intrinsic galactic scaling relationship (e.g., the \citealt{Greene20} $M_\bullet-\sigma$ relation) then one {\it should not use that form of the scaling relationship to infer TDE black hole masses}. This (potentially counterintuitive) statement follows from the fact that if a given (e.g.,) $M_\bullet-\sigma$ law is followed by the intrinsic TDE population, it will not be followed by any given observed TDE population (unless we can be certain we have observed every TDE in a given mass bin). If one wants to use $M_\bullet-\sigma$ to infer TDE host masses, then one must take into account the (empirical) scaling between $L$ and $M_\bullet$, and explicitly calculate which $M_\bullet-\sigma$ relationship {\it observed} TDEs will follow (for a given survey). This biased relationship should then be used for mass inference. 

\subsection{Population synthesis and LSST}
The above calculation was intended to highlight a simple point of statistics (the Malmquist-Hills bias), it also highlights another potential use of {\tt TDEFLARE}, however. With suitable priors on the population of TDE (i) black hole masses (the intrinsic mass function would be a good start), (ii) flare temperatures (a uniform distribution in $\log_{10} T$ between 4 and 5 seems plausible) and (iii) rise timescales (a weak power-law dependence on black hole mass is suggested by the data \citealt{Mummery_et_al_2024}) and a flux limit, one can very simply use {\tt TDEFLARE} to generate a population of TDE light curves. This may be of use for predicting how many TDEs may appear in a future optical survey, or planning optimal follow up of future TDE systems.

\subsection{Can I measure anything else?}
In principle, but only in practice at high black hole masses. The reason for this is that a high black hole mass, in a Bayesian framework, is additional information beyond the priors which can be used to place constraints on the spin of the black hole and the mass of the star which was disrupted. The physics behind this statement is again the Hills mass, which preferentially favors high spins and high stellar masses when the black hole mass is high (comparable to the Hills mass). 

\begin{figure*}
    \centering
    \includegraphics[width=0.45\linewidth]{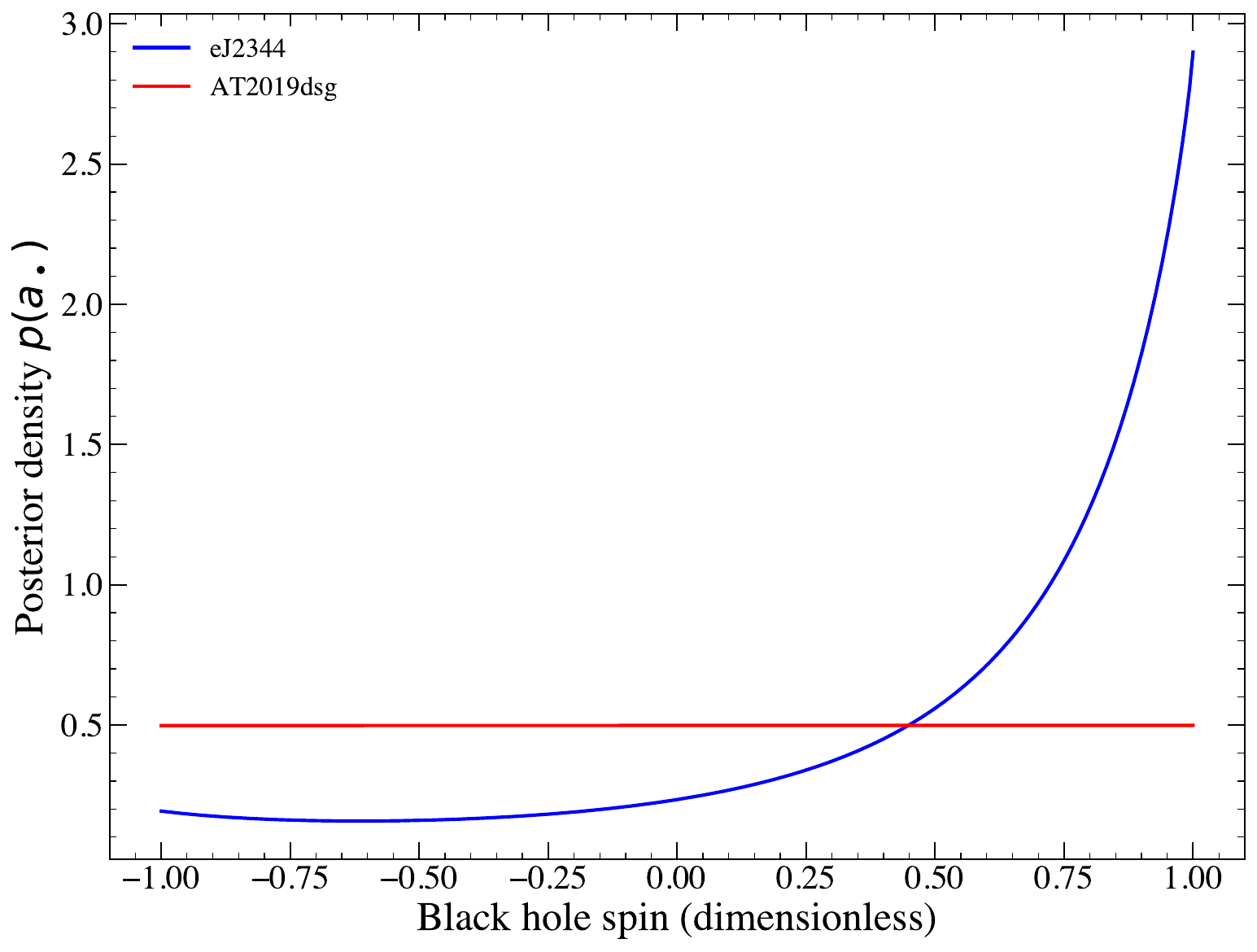}
    \includegraphics[width=0.45\linewidth]{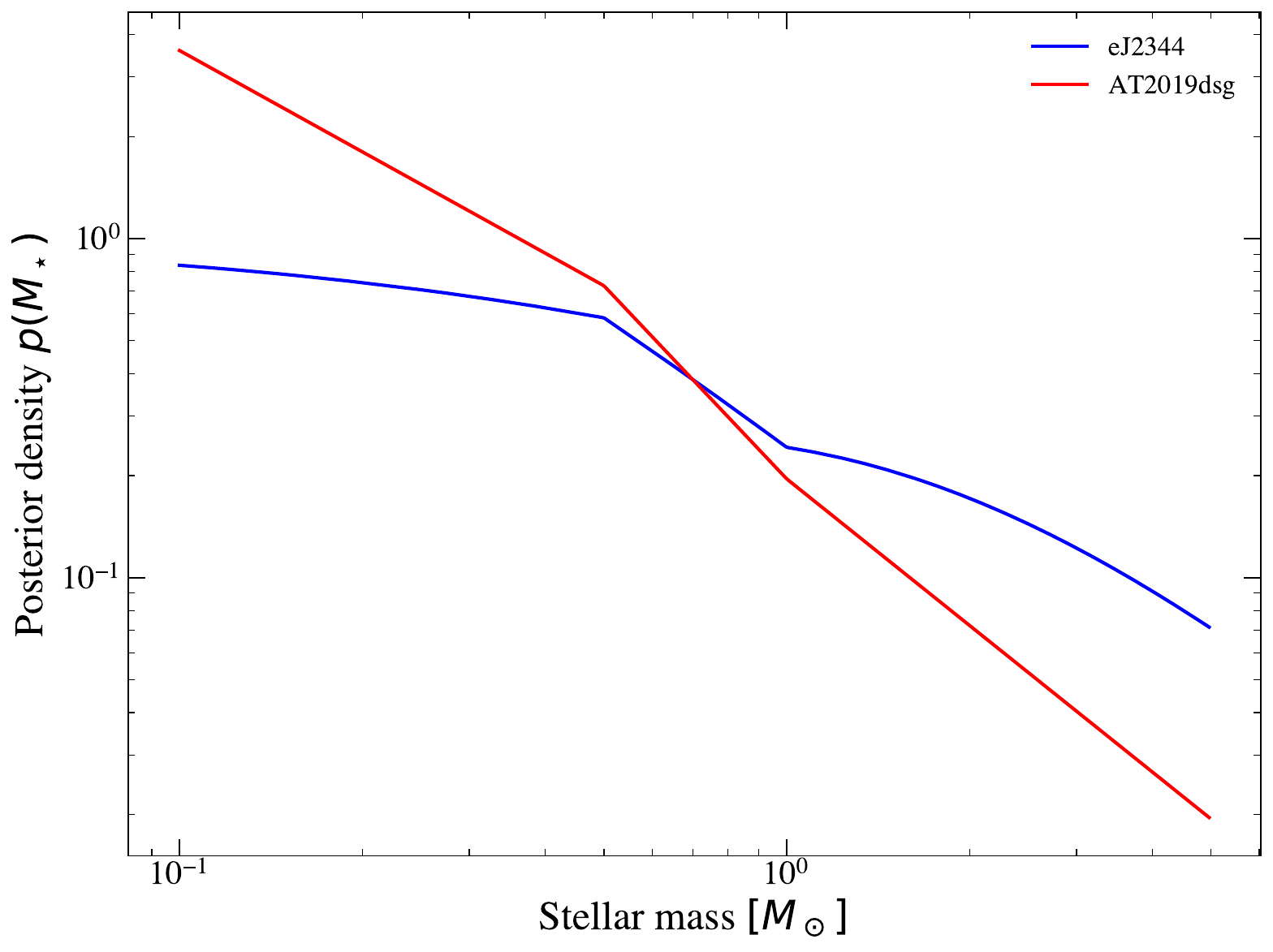}
    \caption{Posterior distributions for the black hole spin (left) and incoming stellar mass (right) can be significantly different from the priors (red curves, returned for the low-mass TDE AT2019dsg) when the black hole mass is comparable to the Hills mass scale. See the blue curves, shown for the (high mass) TDE eJ2344.  }
    \label{fig:spin_and_star}
\end{figure*}

In practice, the posterior distribution of the stellar mass and black hole spin  can be inferred in an analogous manner to the black hole mass, namely
\begin{multline}
    p(a_\bullet|%L_{\rm pk}, 
    E_g, L_P) = p(a_\bullet)\iiint p(M_\bullet|%L_{\rm pk}, 
    E_g, L_P)\,\\ {\cal H}(M_\bullet, M_\star, a_\bullet, \psi)  p(M_\star)  p(\psi) \, {\rm d}M_\star\, {\rm d}M_\bullet \, {\rm d}\psi ,
\end{multline}
and 
\begin{multline}
    p(M_\star|%L_{\rm pk}, 
    E_g, L_P) = p(M_\star)\iiint p(M_\bullet|%L_{\rm pk}, 
    E_g, L_P)\,\\ {\cal H}(M_\bullet, M_\star, a_\bullet, \psi)  p(a_\bullet)  p(\psi) \, {\rm d}a_\bullet\, {\rm d}M_\bullet \, {\rm d}\psi .
\end{multline}
In these expression the pre-factors are the priors (we again use the standard priors in {\tt tidalspin}). For low black hole masses (compared to the Hills mass), the Hills suppression factors ${\cal H}$ is unity for almost the entire range of the integrals, and one simply recovers the priors (red curves in Figure \ref{fig:spin_and_star}, shown for the TDE AT2019dsg), but for masses close to the Hills scale, ${\cal H}$ may be zero for a significant range of the integral domain. 

In this later case (e.g., blue curves in Figure \ref{fig:spin_and_star}, shown for the high-mass TDE eJ2344) the posterior differs significantly from the prior, and a statistical statement about the stellar mass/black hole spin can be made (which is more interesting than the prior). 

\section{Conclusions}\label{conclusions}
The purpose of this work was to provide a practical tool for constraining black hole masses of TDE hosts from the optical/UV data observed from these events. This will become a problem of ever increasing importance as we move into the LSST era with potentially 1000's of sources, offering a fantastic opportunity to constrain black hole masses on the population level and answer various questions of fundamental astrophysical importance. 

We believe that the code {\tt TDEFLARE}, presented in this paper, represents the best population-level tool available for constraining black hole masses from the optical/UV emission in TDEs. It is calibrated using only well understood and robust disk physics, utilizing these tight empirical relationships for regions of observational parameter space which are more poorly understood theoretically. 

We have shown that {\tt TDEFLARE} (i) produces black hole mass constraints consistent with disk codes containing far more physics, (ii) reproduces galactic scaling relationships at high ($>5\sigma$) significance, (iii) produces reliable mass estimates for both partial and full disruptions, and (iv)  does not require late time data to derive mass constraints. 

While {\tt TDEFLARE} provides a reliable method of constraining black hole masses from optical/UV data, it must not be used in place of physically motivated disk models when the data which can be fit with these models (late time optical/UV emission and X-ray spectra) is available. The reasons for this are myriad, but include (i) {\tt TDEFLARE} necessarily lacks broader interpretability, something one gets only with physical models,  (ii) only reliable physically motivated models can provide an insight into outliers (which lie far from the empirical relationships, and are therefore interesting), (iii) {\tt TDEFLARE} ignores multi wavelength emission, which is intrinsically interesting and contains physical information about the system, (iv)  only physical disk models can constrain (e.g.,) the accretion rate through the disk, or the size of the accretion flow, both quantities of interest to broader questions, and (v) morally -- one should always use physical models when possible. Indeed, physically motivated disk models, including {\tt FitTeD} \citep{mummery2024fitted} and {\tt kerrSED} \citep{GuoloMum24} are reliable, well tested and should be used when possible.

Further to this point, the purpose of this work was, explicitly,  {\it not} to solve the (much harder) problem of determining what powers the early-time flares observed from TDEs. Indeed, no attempt has been made to interpret the empirical scaling relationships observed across the TDE population. We stress that physically well-understood models will always be preferable to empirical frameworks -- and should a consensus physical model be developed for the early time evolution of TDE flares in the optical/UV, which can be simply fit to data, then {\tt TDEFLARE} will be redundant. Until that day, however, it seems that a purely empirical but reliable framework like {\tt TDEFLARE} is preferable to more readily interpretable but apparently physically-lacking models. 

\section*{Acknowledgments}
A. Mummery is supported by  by the John N. Bahcall Fellowship Fund at the Institute for Advanced Study. The author is grateful to Sjoert van Velzen and Matt Nicholl for comments on a draft version of this paper, and to Muryel Guolo and Adelle Goodwin for encouragement. 

\section*{Code availability}
The {\tt TDEFLARE} code is available as a model within the {\tt FitTeD} disk code package \citep{mummery2024fitted}, which can be downloaded at \href{https://bitbucket.org/fittingtransientswithdiscs/fitted_public/src}{https://bitbucket.org/fittingtransientswithdiscs/fitted\_public/src}. 

\bibliographystyle{aasjournal}
\bibliography{andy}

\appendix

\section{Mass constraints of all tidal disruption events}
In Table \ref{mass_table} we list all of the mass constraints found from using {\tt TDEFLARE} in this work. All masses are quoted in $\log_{10}(M_\odot)$ units, and the mean and standard deviation are provided. Note that sources near to the Hills mass can have uncertainties below the conflation value, this is a result of the Hills mass providing additional information which further narrows down the mass constraint. 

\renewcommand{\arraystretch}{1.}
\footnotesize
\begin{table}
\centering
\begin{tabular}{ |c|c|c|c| }% Note, generated in python. 
\hline
TDE Name & $M_\bullet$ & TDE Name & $M_\bullet$  \\
\hline
- & $ M_\odot$ & - & $ M_\odot$   \\
\hline
 SDSS-TDE1 & $6.77^{+0.30}_{-0.30}$ &   SDSS-TDE2 & $7.51^{+0.24}_{-0.24}$  \\  
 GALEX-D23H-1 & $6.40^{+0.38}_{-0.38}$ &   PS1-10jh & $6.50^{+0.69}_{-0.69}$  \\  
 AT2017eqx & $6.94^{+0.33}_{-0.33}$ &   PTF-09ge & $6.58^{+0.36}_{-0.36}$  \\  
 PTF-09axc & $6.07^{+0.44}_{-0.44}$ &   PTF-09djl & $6.76^{+0.42}_{-0.42}$  \\  
 ASASSN-14ae & $6.95^{+0.50}_{-0.50}$ &   ASASSN-14li & $5.87^{+0.29}_{-0.29}$  \\  
 ASASSN-15oi & $6.00^{+0.29}_{-0.29}$ &   ASASSN-15lh & $7.71^{+0.21}_{-0.21}$  \\  
 AT2018dyb & $6.82^{+0.29}_{-0.29}$ &   AT2018fyk & $7.09^{+0.27}_{-0.27}$  \\  
 AT2019ahk & $7.25^{+0.26}_{-0.26}$ &   iPTF-15af & $6.17^{+0.34}_{-0.34}$  \\  
 iPTF-16axa & $7.16^{+0.27}_{-0.27}$ &   iPTF-16fnl & $5.78^{+0.29}_{-0.29}$  \\  
 OGLE16aaa & $7.25^{+0.31}_{-0.31}$ &   AT2018zr & $6.83^{+0.29}_{-0.29}$  \\  
 AT2018bsi & $6.27^{+0.31}_{-0.31}$ &   AT2018hco & $7.38^{+0.24}_{-0.24}$  \\  
 AT2018iih & $7.94^{+0.24}_{-0.24}$ &   AT2018hyz & $6.89^{+0.29}_{-0.29}$  \\  
 AT2018lni & $7.31^{+0.25}_{-0.25}$ &   AT2018lna & $6.68^{+0.29}_{-0.29}$  \\  
 AT2019cho & $7.01^{+0.31}_{-0.31}$ &   AT2019bhf & $7.05^{+0.28}_{-0.28}$  \\  
 AT2019azh & $6.59^{+0.29}_{-0.29}$ &   AT2019dsg & $6.83^{+0.29}_{-0.29}$  \\  
 AT2019ehz & $6.74^{+0.29}_{-0.29}$ &   AT2019eve & $6.60^{+0.29}_{-0.29}$  \\  
 AT2019mha & $5.25^{+0.33}_{-0.33}$ &   AT2019meg & $6.49^{+0.37}_{-0.37}$  \\  
 AT2019lwu & $5.66^{+0.35}_{-0.35}$ &   AT2019qiz & $5.99^{+0.29}_{-0.29}$  \\  
 AT2020neh & $6.03^{+0.29}_{-0.29}$ &   AT2019teq & $4.87^{+0.30}_{-0.30}$  \\  
 AT2020pj & $5.07^{+0.32}_{-0.32}$ &   AT2019vcb & $5.71^{+0.35}_{-0.35}$  \\  
 AT2020ddv & $6.13^{+0.37}_{-0.37}$ &   AT2020ocn & $4.72^{+0.30}_{-0.30}$  \\  
 AT2020mbq & $6.00^{+0.37}_{-0.37}$ &   AT2020mot & $7.15^{+0.27}_{-0.27}$  \\  
 AT2020opy & $7.46^{+0.23}_{-0.23}$ &   AT2020zso & $6.04^{+0.29}_{-0.29}$  \\  
 AT2020qhs & $8.17^{+0.24}_{-0.24}$ &   AT2020ysg & $8.07^{+0.24}_{-0.24}$  \\  
 AT2020wey & $5.37^{+0.29}_{-0.29}$ &   AT2020riz & $7.87^{+0.23}_{-0.23}$  \\  
 AT2020vwl & $6.19^{+0.30}_{-0.30}$ &   AT2020acka & $8.35^{+0.22}_{-0.22}$  \\  
 AT2021ack & $6.73^{+0.29}_{-0.29}$ &   AT2021crk & $6.91^{+0.29}_{-0.29}$  \\  
 AT2021axu & $7.50^{+0.22}_{-0.22}$ &   AT2021ehb & $6.31^{+0.33}_{-0.33}$  \\  
 AT2021jsg & $6.50^{+0.37}_{-0.37}$ &   AT2021gje & $8.00^{+0.24}_{-0.24}$  \\  
 AT2021jjm & $6.18^{+0.37}_{-0.37}$ &   AT2021nwa & $6.58^{+0.29}_{-0.29}$  \\  
 AT2021mhg & $6.61^{+0.29}_{-0.29}$ &   AT2021sdu & $6.53^{+0.29}_{-0.29}$  \\  
 AT2021uvz & $6.33^{+0.36}_{-0.36}$ &   AT2021uqv & $7.25^{+0.26}_{-0.26}$  \\  
 AT2021yzv & $7.93^{+0.24}_{-0.24}$ &   AT2021yte & $6.19^{+0.29}_{-0.29}$  \\  
 AT2021lo & $7.65^{+0.21}_{-0.21}$ &   AT2022dbl & $5.93^{+0.25}_{-0.25}$  \\  
 AT2022bdw & $6.21^{+0.30}_{-0.30}$ &   AT2022rz & $7.01^{+0.28}_{-0.28}$  \\  
 AT2022exr & $5.42^{+0.31}_{-0.31}$ &   AT2022dyt & $6.32^{+0.29}_{-0.29}$  \\  
 AT2022dsb & $6.16^{+0.29}_{-0.29}$ &   AT2022hvp & $7.63^{+0.21}_{-0.21}$  \\  
 AT2022pna & $6.70^{+0.36}_{-0.36}$ &   AT2022wtn & $6.99^{+0.32}_{-0.32}$  \\  
 AT2023clx & $5.21^{+0.29}_{-0.29}$ &   AT2022lri & $5.77^{+0.30}_{-0.30}$  \\  
 AT2023rvb & $6.34^{+0.42}_{-0.42}$ &   AT2023cvb & $7.48^{+0.22}_{-0.22}$  \\  
 AT2023mfm & $7.02^{+0.35}_{-0.35}$ &   AT2020yue & $7.75^{+0.22}_{-0.22}$  \\  
 AT2021utq & $6.59^{+0.38}_{-0.38}$ &   AT2020abri & $6.98^{+0.35}_{-0.35}$  \\  
 AT2018jbv & $8.27^{+0.23}_{-0.23}$ &   AT2019cmw & $7.98^{+0.27}_{-0.27}$  \\  
 AT2020ksf & $7.08^{+0.28}_{-0.28}$ &   AT2020vdq & $6.21^{+0.29}_{-0.29}$  \\  
 eJ2344 & $7.83^{+0.21}_{-0.21}$ &  &  \\  
 \hline
\end{tabular}
\caption{The black hole mass constraints of the 89 TDEs in our sample. The quoted error ranges correspond to $1\sigma$ uncertainties.   }
\label{mass_table}
\end{table}

\end{document}